\documentclass{jfm}
\usepackage{graphicx}
\usepackage{epstopdf, epsfig}

\usepackage{xcolor}

\usepackage{newtxtext}
\usepackage{newtxmath}
\usepackage{natbib}
\usepackage{hyperref}
\hypersetup{
    colorlinks = true,
    urlcolor   = blue,
    citecolor  = black,
}

\newcommand{\RomanNumeralCaps}[1]
\linenumbers

\shorttitle{Wet granular flows over a rough incline}
\shortauthor{S. Deboeuf  and A. Fall}

\title{Unsaturated wet granular flows over a rough incline: frictional and cohesive rheology}

\author{Stéphanie Deboeuf\aff{1}
  \corresp{\email{sdeboeuf@dalembert.upmc.fr}}
 \and Abdoulaye Fall\aff{2} 
 \corresp{\email{abdoulaye.fall@univ-eiffel.fr}}
 }

\affiliation{\aff{1} Sorbonne Université, CNRS, UMR 7190, Institut Jean Le Rond d’Alembert, F-75005 Paris, France 
\aff{2} Laboratoire Navier (UMR 8205), CNRS, Ecole des Ponts ParisTech, Univ. Gustav Eiffel Cité Descartes-77420 Champs sur Marne, France 
}

\begin{document}

\maketitle

\begin{abstract}
Multi-phase flows encountered in nature or in industry, exhibit non trivial rheological properties, that can be understood better thanks to model materials and appropriate rheometers. Here, we use model unsaturated granular materials: assemblies of frictional spherical particles bonded by a small quantity of a wetting liquid, 
over a rough inclined plane. 
Our results show steady uniform flows  
for a wide range of parameters (the inclination angle and the mass flow-rate).  
A theoretical model, based on the Mohr-Coulomb yield criterion extended to inertial flows: $\tau=\tau_c + \mu(I) P$,  
in which $\tau_c$ and $\mu(I)$ are the cohesion stress and the internal friction coefficient respectively, gives predictions in quantitative agreement with experimental measurements only when one considers  that dry and wet samples have straightforwardly different internal friction commonly described by the so-called $\mu(I)$-rheology.  
The liquid bridges bounding grains not only induce cohesion, but modify the internal friction 
of the wet assemblies.
\end{abstract}


\section{Introduction}
\label{intro}

Unsaturated wet granular media are triphasic systems composed of an assembly of grains whose pore space is partially filled by a liquid 
and a gas.  
This description applies to numerous materials in civil engineering. They exhibit non common behaviours and their rheological characterization is an active research topic due to the prevalence of such flows in both natural (landslides, debris flows, mudflows, rock and snow avalanches) and industrial fields (many powder-processing methods such as wet granulation, fertilizer production or coating). Our understanding of their flow behaviour, however, is still fragmental due to the system complexity, which implies a difficult experimental instrumentation and lack of unified theoretical approaches. The influence of a liquid wetting the grains on the granular material rheology sensitively depends on the liquid content (or saturation) and morphology of the liquid phase~\citep{Mitarai06,Scheel08}. 
It is commonly known that the addition of a small amount of liquid in a granular medium creates cohesion properties due to the surface tension of the wetting liquid, which forms pendular bridges attracting grains to their close neighbours. Such a mixture of grains, liquid and air may have a strong solid-like behaviour~\citep{Strauch12} that enables the building of sand castles as opposed to dry sand which cannot stabilize under gravity with slopes steeper than the angle of repose~\citep{Fiscina12,Bocquet98}.
Due to the very high solid fraction of grains in such materials,  
direct inter-grains contacts play an important part in the physical mechanisms ruling their rheological behaviour, which is strongly influenced by the shape, surface and material properties of the grains (roughness, friction, elasticity, ...). To those micromechanical features, one should add the physical characteristics of the liquid (its viscosity and surface tension, ...) and geometrical microstructural information\citep{Mitarai12,Scheel08,Badetti18}.

Starting from what is done for a frictional granular flow~\citep{GDRMiDi04,DaCruz05,Andreotti13}, 
dimensional analysis of a frictional cohesive granular flow described by the shear, normal and cohesion stresses   $\tau$, $P$ and $\tau_c$, the shear rate $\dot{\gamma}$ and the material properties $\rho$ and $d$, the mass density and grain diameter respectively, leads to the identification of three dimensionless parameters,  defined as: 
\begin{eqnarray}
\mu^{eff} &=& \frac{\tau}{P} \\
\mu_c &=& \frac{\tau_c}{P} \\
I &=& \frac{\dot{\gamma} d}{\sqrt{P/\rho}} \label{I}
\end{eqnarray}
and the rheological law, formulated as: 
\begin{eqnarray}
\mu^{eff} ( \mu_c ,I). \label{mueffwet00}
\end{eqnarray}
Liquid partially filling pores between grains is supposed to induce some cohesion, taken into account by the  effective stress cohesion $\tau_c$, that appears within the cohesion ratio $\mu_c$. In addition, one retrieves the  effective friction $\mu^{eff}$ and the inertial number $I$ common to dry (cohesionless) frictional granular flows~\citep{GDRMiDi04,DaCruz05,Andreotti13}. Besides, one retrieves the frictional law $\mu^{eff} (I)$ if $\mu_c=0$ in the absence of cohesion. 

More knowledge about this cohesive frictional law $\mu^{eff} ( \mu_c ,I)$ can be deduced from 
 traditional approaches in solid mechanics, often used in geomechanics~\citep{Pierrat98,Cleaver00,Mitchell05} and extended to static partially saturated granular materials, namely the Mohr-Coulomb yield criterion. 
  On the macroscopic scale, the effect of cohesion is described in the quasi-static limit of slow flow by:
\begin{eqnarray}
\tau &=& \tau_c + \mu\, P,  \label{eq_rheocohesionstat}
\end{eqnarray}
where $\tau_c$ is the macroscopic cohesion that vanishes for cohesionless materials and $\mu$ the macroscopic friction coefficient of the dry
grain assemblies~\citep{Pierrat98,Badetti18b}. We will see later that $\mu$  should be rather the internal friction coefficient of the wet grain assemblies.  
Within such a formulation, cohesion and internal friction are taken into account by separate contributions. This can be formulated as: 
\begin{eqnarray}
\mu^{eff}  &=& \mu_c + \mu, \label{mueffwet0}
\end{eqnarray}
where $\mu^{eff}=\tau / P$ is the effective friction of the wet grain assemblies, while $\mu$ is the internal friction.   
Can this plastic flow rule be extended to non-quasi static but inertial cohesive granular flows? 

For cohesionless dry granular materials in quasi-static and inertial shear flows, the inertial number $I$ defined in equation~(\ref{I}) 
 has been adopted over the past 20 years~\citep{GDRMiDi04,DaCruz05,Andreotti13} as a relevant parameter describing their internal state through the macroscopic 
 friction $\mu^{eff}$ so that: 
\begin{eqnarray}
\tau &=& \mu^{eff} (I)\, P. 
\end{eqnarray}
In this particular case, equation~(\ref{eq_rheocohesionstat}) still holds by considering $\tau_c=0$  both for quasi-static and inertial flows: internal and effective frictions $\mu$ and $\mu^{eff}$  are equal.
Whereas it is known for a while that the friction coefficient $\mu=\mu^{eff}$  is an increasing function of the inertial number $I$ for cohesionless granular materials~\citep{GDRMiDi04,Staron10,Fall15,Saingier16}, this feature has been studied and discussed in a number of experimental and numerical recent publications~\citep{Rognon08,Richefeu08,Badetti18,Badetti18b,Khamseh15,Berger16,Vo20, Mandal8366} for cohesive granular materials. 

For cohesive wet grains, it has been shown~\citep{Rognon08,Khamseh15,Berger16,Badetti18,Badetti18b} that the rheology can be described by two dimensionless numbers: the inertial number $I$
 and the reduced pressure $P^*=Pd^2/F_0$ which compares confining forces $Pd^2$  to cohesive forces (at the grain scale) $F_0$, that can be approximated as $F_0\simeq\upi\Gamma d$ 
for grains of diameter $d$ joined by menisci with surface tension $\Gamma$~\citep{Raux18}. 
Wet granular materials exhibit a similar behaviour to dry ones, with an additional dependence on the reduced pressure $P^*$: the effective friction coefficient  $\mu^{eff}$ increases when $P^*$ decreases, and the solid volume fraction $\phi$ grows with $P^*$~\citep{Berger16,Badetti18b}.
Very recently, \citet{Vo20} proposed that both numbers should be combined into one single visco-cohesive-inertial number which could solely control macroscopic stresses and micromorphologies.

Note that, the dimensionless parameter $P^*$ defined above needs the knowledge a priori of the cohesive forces at the microscopic scale.  
Alternatively, we prefer here to introduce the dimensionless pressure $P/\tau_c=1/\mu_c$ to avoid confusion with the definition of $P^*$ from the literature. Indeed, to the aim to solve continuous materials mechanics equations, one needs to identify the relevant rheological law at a macroscopic scale. 
 
A standard geometry to study the flow of granular materials is the inclined plane. Indeed, the free-surface flow of particulate solids down an incline is an interesting example of a granular flow that not only has a number of engineering applications but also takes place in numerous types of natural phenomena. Furthermore, for theoreticians concerned with constitutive modelling, the inclined plane flow can serve as a viscosimetric flow, since regimes of steady uniform flows lead to effective friction controlled experiments~\citep{GDRMiDi04,Savage83,Savage84,Pouliquen96,Saingier16}.
Surprizingly, among the broad literature about the free-surface flow of granular materials down an incline, only a few studies~\citep{Brewster05} were dedicated to the case of cohesive materials and the experimental constitutive law in term of $\mu(I)$ or $\mu^{eff}(I)$ rheology has not yet been reported. 
Our general questions are: Which rheology would apply to a gravitational multi-phase flow seen as a continuous fluid?
The rheology we will use here is formulated as:
\begin{eqnarray}
\tau &=& \tau_c + \mu(I) P. 
\label{eq_rheocohesion}
\end{eqnarray}
Do predictions with the rheology from equation~(\ref{eq_rheocohesion}) systematically compare with our experiments,  realized for larger systems and larger values of the inertial number $I$ classically studied? Recently, some predictions with this rheology have been computed by~\citet{Abramian20} for the stability of a cohesive granular column and compared with numerical discrete simulations.

To this aim, we report in the present article  an experimental study of the flow of frictional, cohesive grains down a rough inclined plane.  Cohesion is induced here by liquid partially saturating a granular material.  
We will compare such  free-surface gravitational flows of wet grains with flows of dry grains. 
First, the experimental setup and the measurement procedures are outlined in Section~\ref{methods}, followed by our  theoretical predictions, based on the choice of the internal friction law in Section~\ref{theo}. Then the main results of this study are summarized in Section~\ref{results}. Several discussions about different rheological models for cohesive granular flows 
are given in Section~\ref{disc}, before conclusions are drawn in Section~\ref{conclu}.

\section{Experimental methods}
\label{methods}

\subsection{Experimental setup and materials}

Cohesion and the dynamic behavior of wet granular materials can display a non trivial dependence on the amount and type of the wetting liquid~\citep{Raux18}.  
In this work, we focus on the pendular state~\citep{Mitarai12}, in which liquid bridges are small~\citep{Badetti18,Scheel08}.
The experiments are carried out on model materials: an assembly of sub-macroscopic solid spherical beads, mixed with a nonvolatile, wetting, Newtonian liquid. We use rigid glass beads of density $\rho=2500$kg.m$^{-3}$ and of diameter $d\in[200$-$300]\mu$m $\simeq250\mu$m and the wetting liquid is a silicon oil  (47V20) of viscosity $\eta=20$mPa.s  and of surface tension $\Gamma=20.4$mN.m$^{-1}$. The surface tension is measured with an automatic single drop tensiometer (TRACKERTM, by Teclis Scientific). The Bond number of our cohesive material, comparing cohesive to gravitational forces, is thus small enough to be in a cohesive regime: $ Bo=\rho g d^2 / (\upi \Gamma) = 10^{-2}\ll1$.  

We always prepare the system for measurements in the same way to ensure reproducible experimental conditions: the wetting liquid is mixed thoroughly with the dry beads until a uniform consistency is visually obtained. The liquid content $\epsilon$ defined as the mass ratio of liquid over grains is constant $\epsilon=0.5\%$. The solid fraction is supposed to be equal to $\phi\approx0.6$, by default of more precise measurements. We will come back later to this assumption, that has only few effects on our results.

\begin{figure} 
	\centering
	\begin{minipage}{.3\linewidth} 
	\includegraphics[width=\linewidth]{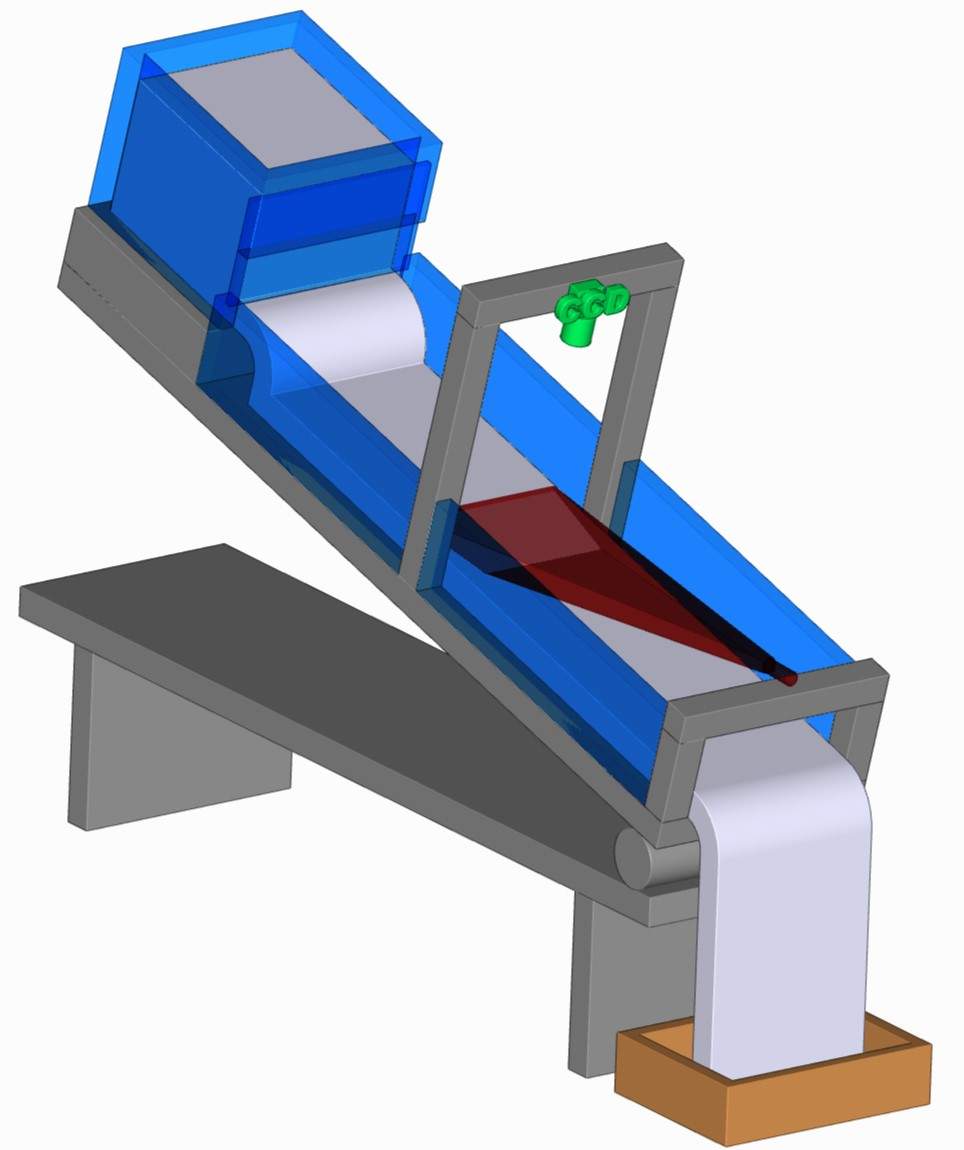} a)
	\end{minipage} 
	\begin{minipage}{.3\linewidth}
	\includegraphics[width=\linewidth]{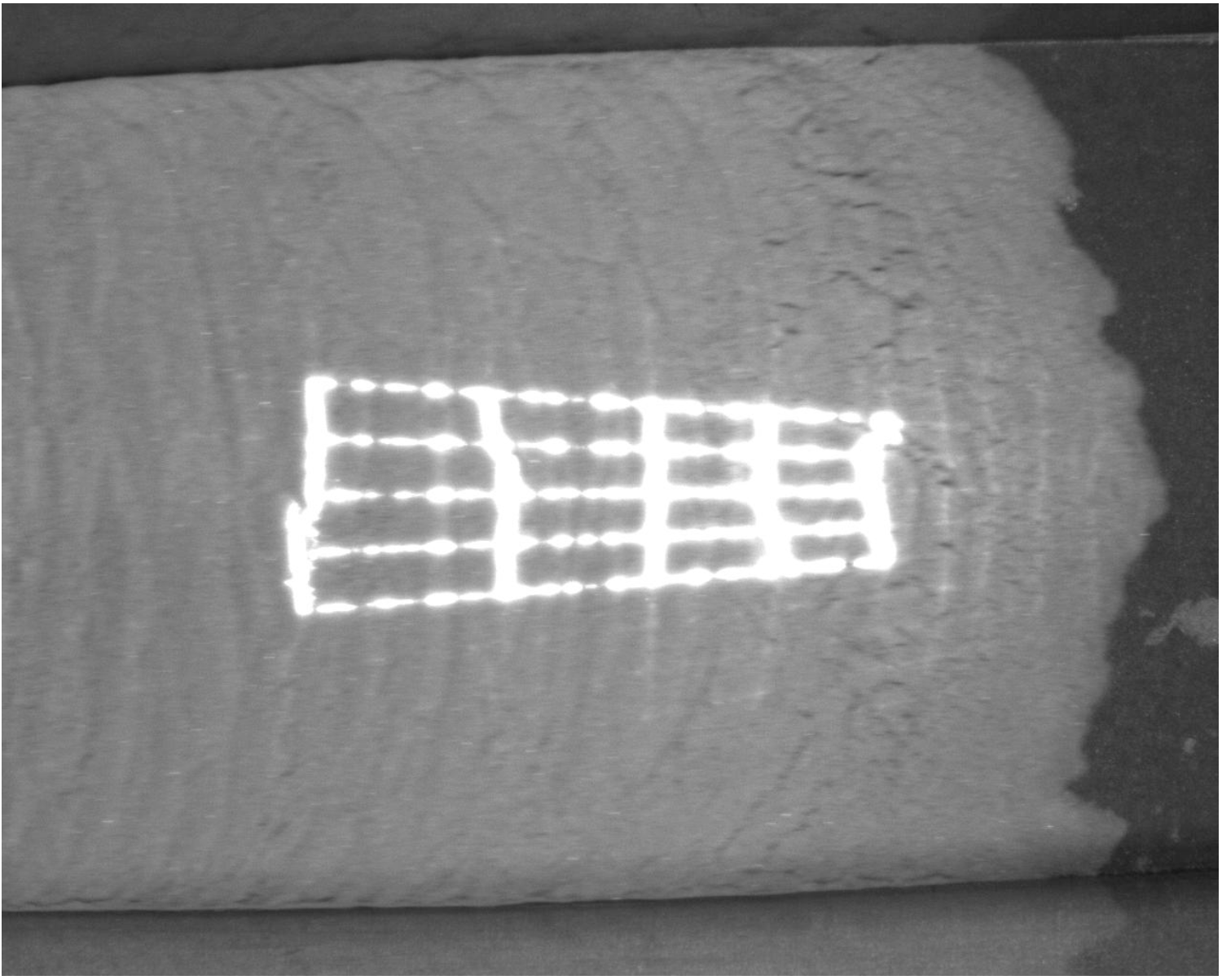} b)
	\end{minipage} 
	\caption{a) The experimental setup consists of a rough plane that can be inclined at the imposed slope angle $\theta$. Wet or dry grains, stored in a hopper at the top of the plane, are released through a gate, which aperture thickness $f$ can be adjusted in order to control the mass flow-rate $Q_m$, measured by means of a scale located underneath. The granular layer free surface is observed thanks to a CCD camera. A LASER diode is used to produce a grid or line sheet that makes possible to do profilometry and measure  height profiles $h(x,t)$. b) A picture of the free surface of a wet granular flow on the incline at $\theta=31^{\circ}$ and $f=5$cm. 	}
	\label{fig_disp}
\end{figure}

A mass of about $20$kg of (wet or dry) grains is poured  into the hopper located at the top, before opening its frontal gate to start their flow over the inclined plane (figure~\ref{fig_disp}a). 
 The  $1.6$m-long incline is made of a rough bottom plate and two smooth lateral Plexiglass walls. The rough surface is made of sandpaper of roughness $\simeq350\mu$m, which is approximately the same size than the flowing beads (to ensure a no-slip boundary condition at the bottom). The width between the side walls is fixed to $W=34$cm, large enough to neglect sidewall effects~\citep{Jop05}. This hopper, from which particles are released, is elevated with respect to the incline. Thus, when opening the  hopper's gate, the beads first fall down continuously on the rough plane before flowing over it. This thin and dilute ‘rain’ allows a good reproducible initial condition. 
 For high enough flow-rates and slopes, a steady state regime then develops with a finite and (quasi-)constant thickness layer of (wet or dry) grains throughout the rough plate. Note that the free surface of the densely packed wet grains in the hopper is discontinuous with localized and intermittent sliding planes by contrast to dry grains that show a smooth free surface.

In addition, the dense wet grains flowing over the rough plane look like a continuous material with a slightly curved shape with some small waves (figure~\ref{fig_disp}b). Just behind this interface, the wet granular layer is not uniform but exhibits some front profile, and propagates at a constant velocity denoted as $u_0$, so this part of the flow is not uniform but steady. Far away from the front interface, the thickness of the wet granular layer seems roughly uniform and steady, denoted as $h$. A finer inspection shows small surface waves appearing on the wet granular layer (figure~\ref{fig_disp}b), that are not observed for a dry granular flow. 

Finally, upon the quick closing of the hopper gate, the flow rate progressively decreases, leading to a final deposit on the incline. The wet deposit morphology exhibits some characteristics that are not observed for the dry case, such as coexistence of different typical length-scales related to the surface waves present during the flow and to fractures of big clusters of wet grains. The scope of this paper is the study of a steady and uniform wet granular flow over an incline by performing experimental measurements and theoretical computations from the rheology in equation~(\ref{eq_rheocohesion}).  The other observed phenomenologies will be treated in future publications.
The controlled parameters are the gate's aperture height $f$ that varies between  $0.5$cm and $6$cm and the incline angle $\theta$ which is between $24^{\circ}$ and $39^{\circ}$. We will see below that the flow-rate is thus controlled by the opening of the gate at height $f$.

\subsection{Experimental measurements}
\begin{figure}
	\centering
 	 a) \includegraphics[width=.45\linewidth]{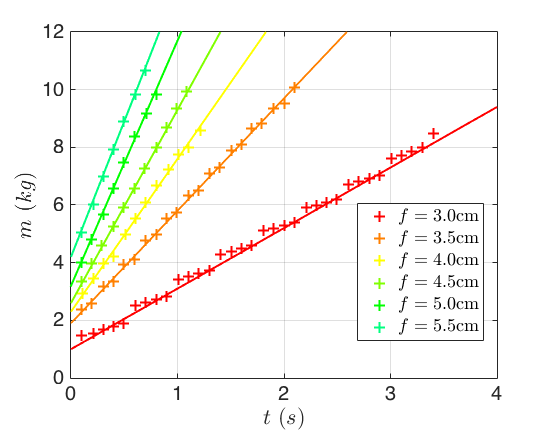}
  	 b) \includegraphics[width=.45\linewidth]{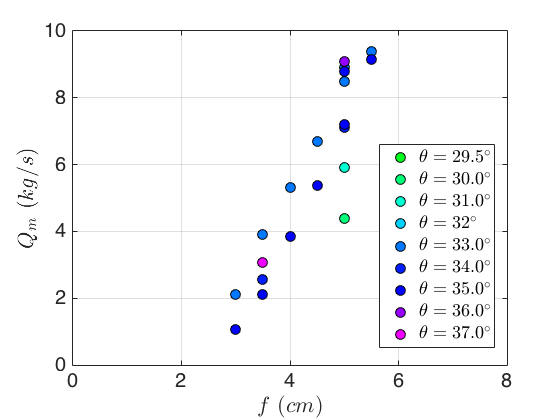} 	  
	\caption{  a) Temporal evolution of the mass  $m(t)$ of wet grains released out of the incline as a function of time for a series of experiments at a constant slope angle $\theta=33^{\circ}$ and for different aperture heights $f\in[3.0$-$5.5]$cm. b) The mass flow rate $Q_m(f)$ as a function of  the hopper aperture height $f$ for different incline slope angles~$\theta\in[29.5$-$37.0]^{\circ}$.}
	\label{fig_massflowrate}
\end{figure}

To measure the flow-rate $Q_m$, the mass $m$  of 
wet grains flowing out of the incline is weighted as a function of time. The time evolution of the mass $m(t)$ is drawn in figure~\ref{fig_massflowrate}a for a given inclination angle $\theta$ and different values of the gate height $f$. It is found that $m(t)$  is proportional to time and thus a constant flow-rate $Q_m$ can be defined. In the whole experimental range of  $f$ and $\theta$ explored here, the measurements of $Q_m$ are reproducible and the flow-rate $Q_m$  mainly depends on the hopper gate aperture  $f$, while it is roughly independent of $\theta$ as shown in figure~\ref{fig_massflowrate}b.

\begin{figure}
	\centering
 	 a) \includegraphics[height=.35\linewidth]{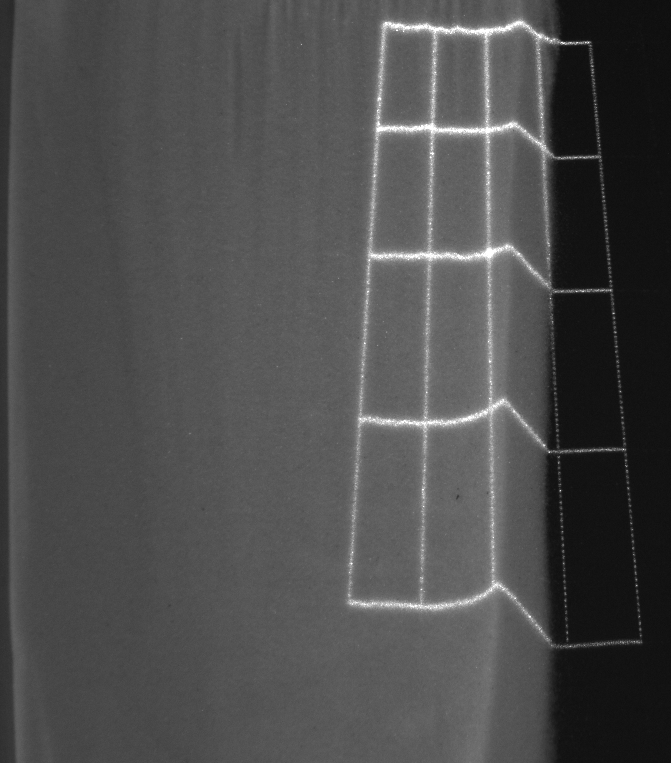}
	 b) \includegraphics[height=.37\linewidth]{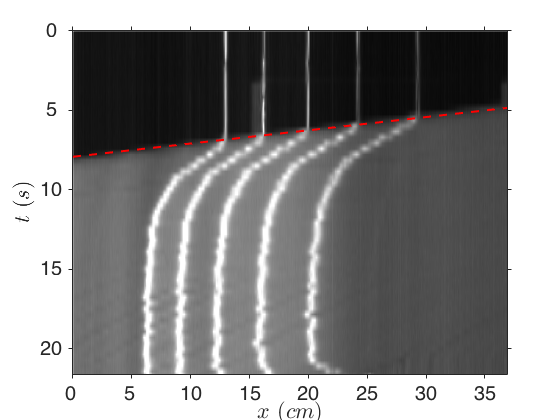} 
	\caption{ a) A picture of the LASER grid pattern showing the flow border giving a measure of $h$. b) Spatial-time diagram showing the juxtaposition of a line of pixels for successive time steps $t$ for $\theta=29.5^{\circ}$  and $f=3.5$cm. This line of pixels is taken from a picture of the flow surface: it is longitudinal to the incline (along $x$) and crosses the flow front in its center. The red dashed straight line shows the front propagation at a constant velocity. }
	\label{fig_hmeas}
\end{figure}

All the measurements presented in the following are done in the steady regime, far from the entrance and the exit of the rough inclined plane, in which the flow is (quasi-) uniform along the transverse direction. 
A grid pattern is projected on the plane by a laser sheet, with a projection angle small enough so that the presence of any granular mass on the plane induces a significant deformation of the grid. Then the local shift  observed between the deformed grid pattern and the initial one without grains (figure~\ref{fig_hmeas}a) is proportional to the thickness $h(x,t)$ of the granular layer at the position $x$ along the longitudinal direction and the time $t$~\citep{Pouliquen99}. The direction normal to the plane is denoted $z$. 

The grains in the central region (unaffected by the non-fully developed region near the gate) is recorded by a CCD camera positioned at the normal of the plane  at a frequency $20$Hz. From these movies, the thickness $h(x,t)$  of the steady-uniform flow of the granular layer is computed. The front velocity $u_0$ is obtained by tracking the front propagating  down the incline with a time-space diagram (figure~\ref{fig_hmeas}b) which is the juxtaposition of a line of pixels for successive time steps, taken from a picture of the flow surface, longitudinal to the incline (along $x$) and crossing the flow front in its center. 
Note that the mass balance equation $\partial h/\partial t + \partial(h\overline{u})/\partial x  = 0$ with  $\overline{u}$, the $z$-averaged velocity, written for a 
 front moving at a constant velocity $u_0$ without deformation, leads to a travelling wave 
 $ h(x,t) = h( \xi = x - u_{0}\,t )$,  and gives 
 $ \mathrm{d} (h(\overline{u} - u_{0}))/ \mathrm{d} \xi = 0$, 
 implying that $\overline{u} = u_0$~\citep{Saingier16}. Thus the front velocity $u_0$ and the  velocity $\overline{u}$  averaged over $z$   are equal.  
All movies and images are processed using ImageJ and Matlab.  The same measurements are done for wet and dry granular flows. 

\begin{figure}
	\centering
	 \includegraphics[width=.45\linewidth]{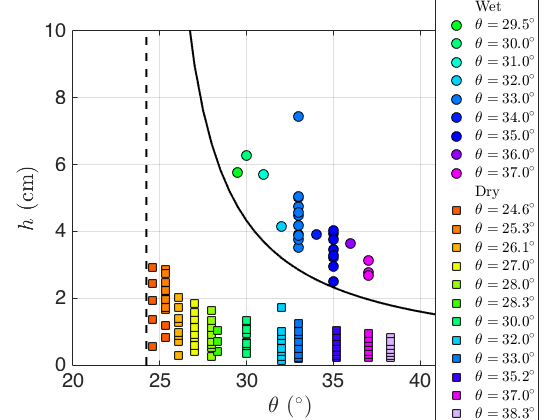} 	
	\caption{ Parameters space ($\theta$, $h$) of the experiments reported in this paper, with $\theta$ the incline slope angle  and $h$ the constant height of the wet (circles) and dry (squares)  granular flows. The same colors encode for the value of $\theta$ both  for wet and dry grains.  Theoretical curve $h_c(\theta)$ computed from equation~(\ref{eq_hc}) with $\tau_c=70$Pa and  $\mu_0=0.45$, for the height threshold for a cohesive steady uniform free-surface gravitational flow. The dashed vertical line corresponds to the angle threshold $\theta_0=24^{\circ}=\arctan(\mu_0)$. }
	\label{fig_param}
\end{figure}

We realized about $30$ experiments of steady-uniform wet granular flows and $100$ experiments of dry granular flows. 
Figure~\ref{fig_param} shows the  thickness $h$ of steady-uniform flows for the different inclinations $\theta$ explored for both the dry (squares) and the wet (circles) granular samples. Steady-uniform flows are achievable for large enough values of $h$ and $\theta$. 
 One can see that wet granular flows are much thicker than dry ones.
Explored data ($\theta$, $h$)  for the wet samples are all above the predicted evolution of the minimal flowing height $h_c(\theta)$ for a cohesive steady uniform free-surface gravitational flow, drawn in black and computed from equation~(\ref{eq_hc}) 
without any fit parameter as soon as some minimal rheological parameters ($\tau_c=70$Pa and  $\mu_0=0.45$) were determined from the experiments as it will be demonstrated below. The dashed vertical line corresponds to the angle threshold $\theta_0=24^{\circ}=\arctan(\mu_0)$. 

We will now compute theoretical predictions of the cohesive and frictional rheology encoded by equation~(\ref{eq_rheocohesion}) for  steady uniform free-surface gravitational flows, allowing for subsequent comparisons and analyses of our experimental results.

\begin{figure}
	\centering
	 \includegraphics[height=.25\linewidth]{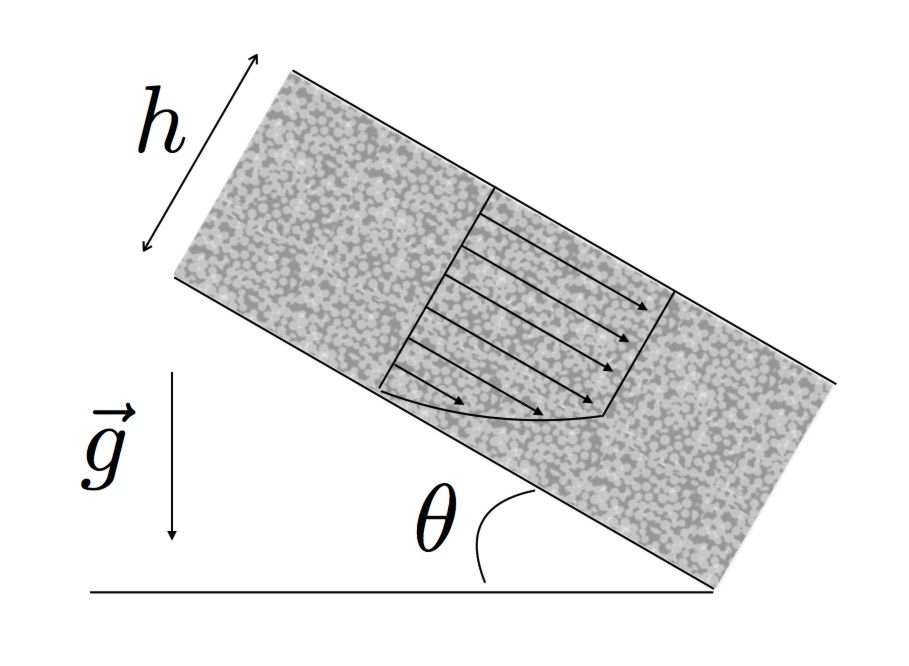}  	 	
	\caption{ Schematic representation of the  vertical section of a wet granular flow. }
	\label{fig_schematheo}
\end{figure}

\section{Theoretical modeling and choice of the rheology}
\label{theo}

From a constitutive modeling point of view, accurate identification and characterization of the rheological laws are in excessive need. Previously, a Mohr-Coulomb yield criterion model has been used for quasi-static and extended to the inertial dense flow regime in cohesive granular materials~\citep{Khamseh15,Berger16,Badetti18,Badetti18b}, written in equation~(\ref{eq_rheocohesion}). We derive here  predictions from this rheology in the case of steady uniform free-surface gravitational flow. 

We use the variable $P$ for the pressure or the normal stress indifferently, considering no normal stress difference here, as often assumed in thin free-surface flows when they are uniform. This may not be the case for three-dimensional configuration with multi-dimensional shear. 

First,  the momentum equations give for a steady fully developed free surface  flow down an incline at an angle $\theta$ (as shown in figure~\ref{fig_schematheo}, the normal and shear stresses,  $P$ and $\tau$ respectively, on a plane parallel to the incline:
\begin{eqnarray}
P &=& \phi \rho g (h-z) \cos\theta,   \label{eq_P} \\
\tau &=& \phi \rho g (h-z) \sin\theta,   \label{eq_tau}
\end{eqnarray}
Thus the inclined plane geometry, whatever the rheology is, in case of a steady uniform flow, allows to  control the macroscopic or effective friction: 
\begin{eqnarray}
\mu^{eff} = \frac{\tau}{P} &=& \tan\theta,  
\end{eqnarray}
throughout the whole layer thickness. The cohesive frictional rheology in equation~(\ref{eq_rheocohesion}) used here  implies that:
\begin{eqnarray}
\mu^{eff} &=&  \frac{\tau_c}{P} +\mu  \label{eq_muwet}
\end{eqnarray}
Contrary to  cohesionless granular materials, the macroscopic friction $\mu^{eff}=\tau/P$ differ from the internal friction $\mu$ for a cohesive granular material.

Thus, to be able to make theoretical computations for wet granular flows from the rheology in equation~(\ref{eq_rheocohesion}), one needs to know the function $\mu(I)$. Different choices are possible:
\begin{itemize}
\item an affine function for $\mu(I)$: 
$\mu(I)=\mu_0 +a \, I $ as in~\citep{DaCruz05}
\item a more elaborate model 
 $\mu(I)= \mu_0 +\Delta\mu/(1+I_0/I)  $ as in~\citep{Jop06}. 
\end{itemize}
Note that the elaborate function is approximated by the affine function for small values of the inertial number $I$. 
First, we will do some general predictions for any function $\mu(I)$ in which $\mu(I=0)=\mu_0$. Second, we will detail computations for the affine function. Then we will show that the affine function and the more elaborate model lead approximately to the same results in the range of our experimental control parameters. Even if both models give analytic relations, the affine law has the advantage of giving compact ones. The relevance of the approximation $\mu(I) = \mu_0$ for wet granular flows is discussed later (section~\ref{disc}). 

Just below, for flows at imposed ($\theta$, $h$), we will compute the normal profiles (normal to the rough inclined plane, along $z$) of the friction coefficient $\mu(z)$ and the inertial number $I(z)$ for given rheological  parameters, that will allow us to deduce the velocity profile $u(z)$, as well as the mass flow-rate $Q_m$ and some measurements achievable in our experiments.

\subsection{Without any a priori  about $\mu(I)$ except $\mu(I=0)=\mu_0$ }

In general, a yield-stress $\tau_Y$  (so that the shear rate $\dot{\gamma}\neq0$ for a shear stress $\tau\geq\tau_Y$) leads to the emergence of a length-scale $h_Y$, as a height threshold for shear in a steady uniform gravitational flow of density $\rho$: $h_Y=\tau_Y/(\rho g\, \sin\theta$).

In a similar way, cohesion (encoded by the stress $\tau_c$ in the frictional Mohr-Coulomb model, so that $\tau\geq\tau_c + \mu_0 P$ if $\dot{\gamma}\neq0$), 
leads to a height threshold $h_c$ for shear in a steady uniform gravitational flow of density $\phi\rho$: 
\begin{equation}
h_c = \frac{\tau_c}{\phi\rho g \cos\theta} \frac{1}{(\tan\theta-\mu_0)} . \label{eq_hc}
\end{equation}
This relation $h_c(\theta)$  is drawn in figure~\ref{fig_param} for $\tau_c=70$Pa and  $\mu_0=0.45$, which values will be justified to be relevant for our wet experiments later in the paper (section~\ref{wetrheo}). As expected, it is below the values  ($\theta$, $h$) of the steady uniform wet granular flows reported in this paper. 

So, for a wet granular layer of height $h\geq h_c$, shear will be localized at the bottom layer of height $h-h_c$, whereas the top layer (of thickness $h_c$) will have no shear but will slide like a plug. At a given cohesion stress $\tau_c$, this critical height has a minimal value equal to $h_c^{min} (\theta=90^{\circ}) = \tau_c / (\phi\rho g)$ that is a few millimeters for $\tau_c$ of the order of $10$Pa: 
$h_c^{min}(\tau_c=70$Pa$) \simeq 5$mm.  
By contrast, in the dry case, there is no  typical length-scale, but only a slope angle threshold $\theta_0=\arctan(\mu_0)$ above which a granular layer can flow steadily and uniformly. 

Only the cohesion stress $\tau_c$ is needed to compute  the local friction coefficient $\mu(z)$ through:
\begin{equation}
 \mu(z) = \tan\theta - \frac{\tau_c}{\phi\rho g (h-z) \cos\theta}.  \label{eq_muz}
\end{equation}
From equation~(\ref{eq_muz}), one checks that the internal friction coefficient $\mu(z)$  depends nor on the function $\mu(I)$  neither on its minimum $\mu_0$. Not only the effective friction $\mu^{eff}$ is controlled here, but $\mu(z)$ too (even if it is not constant with $z$), as soon as the stress cohesion $\tau_c$ is well defined and the rheology of equation~(\ref{eq_rheocohesion}) is relevant. 
 The internal friction coefficient $\mu(z)$ is maximal at the bottom:
 $\mu^{max} =  \mu(z=0) = \tan\theta - \tau_c/(\phi\rho g h \cos\theta)$  
  and decreases for increasing values of $z$; it is equal to $\mu_0$ at the interface $z=h-h_c$ and goes to negative values until $-\infty$ for larger $z\geq h-h_c$. We recall that this internal friction is a material property.  
 For a constant height $h$ (or slope angle $\theta$), the function $\mu(z)$ is larger for larger $\theta$ (or $h$), so that the interface position $h-h_c$ is larger too.

\begin{figure}
	\centering
 	 a) \includegraphics[width=.45\linewidth]{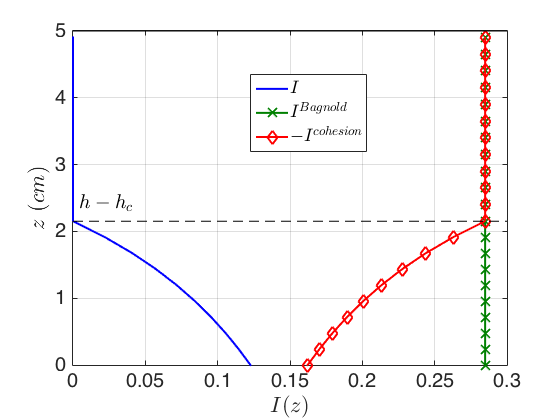} 
	 b) \includegraphics[width=.45\linewidth]{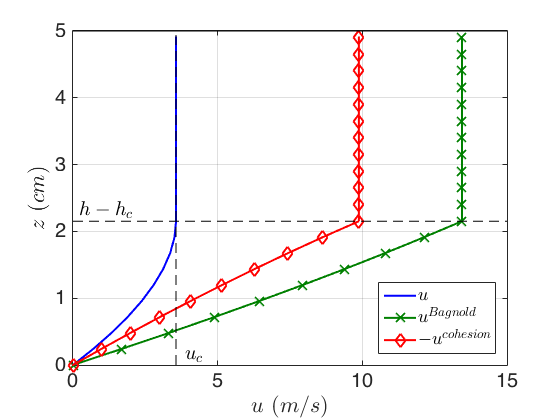} 
	\caption{Theoretical normal profiles of inertial number $I(z)=I^{Bagnold}(z)+I^{cohesion}(z)$  (a)
	and longitudinal velocity $u(z)=u^{Bagnold}(z)+u^{cohesion}(z)$ (b), and their respective Bagnold and cohesion contributions, for one set of parameters ($\theta=33^{\circ}$, $h=5$cm, $\tau_c=70$Pa, $\mu_0=0.45$ and $a=0.7$).}
	\label{fig_contributionBagCoh}
\end{figure}

\subsection{Affine function $\mu(I)$: $\mu(I)=\mu_0+ a \, I$} 
From the affine function $\mu(I)$ and the normal profile $\mu(z)$, the local inertial number $I(z)$ can be computed, by distinguishing two contributions --a cohesion one indicated by the exponent $^{cohesion}$ (equal to $0$ for dry grains when $\tau_c=0$Pa) and a Bagnold one indicated by the exponent $^{Bagnold}$  (common to dry and wet flows in the shear zone)--, as follows: 

\begin{eqnarray}
I(z) &=&  I^{Bagnold}+I^{cohesion}(z) \label{eq_Imuaff}
 \end{eqnarray}
with 
\begin{eqnarray}
 I^{Bagnold} &=& \frac{1}{a} (\tan\theta-\mu_0)  \geq 0  \label{eq_Imuaff2}
 \end{eqnarray}
and 
\begin{eqnarray}
  I^{cohesion}(z) &=& - \frac{1}{a} \frac{\tau_c}{ \phi\rho g (h-z) \cos\theta } =  - I^{Bagnold}  \frac{h_c}{h-z}  \leq 0  \label{eq_Imuaff3}
 \end{eqnarray}
 so that equation~(\ref{eq_Imuaff}) can be written as:
\begin{eqnarray}
I(z) &=& I^{Bagnold} \left( \frac{h-h_c-z}{h-z}\right) \geq 0. \label{eq_Imuaff4}
 \end{eqnarray}
 
 These three profiles are plotted in figure~\ref{fig_contributionBagCoh}a for one set of material properties and flow parameters: $\tau_c=70$Pa, $\mu_0=0.45$, $a=0.7$ and $\theta=33^{\circ}$, $h=5$cm. 
At the interface $z=h-h_c$ between the shear bottom layer and the top plug layer, we have $I^{cohesion}(z=h-h_c) = - I^{Bagnold}$, so that $I(z=h-h_c)=0$.  
At a given $(h$, $\theta)$, the inertial number is maximal at the bottom: 
\begin{eqnarray}
  I^{max} = I(z=0) &=&    
    \frac{1}{a} (\tan\theta-\mu_0) \frac{ h -h_c }{h} , \label{eq_Imax}
 \end{eqnarray}
 and vanishes  at the interface $z=h-h_c$ and above. For a constant $h$ (or $\theta$), the function $I(z)$ is larger for a larger $\theta$ (or $h$). 

Relations~(\ref{eq_Imuaff})-(\ref{eq_Imuaff3}) give the following equation for the shear rate $\dot{\gamma}(z)$ for $z\leq h-h_c$: 
\begin{eqnarray}
\dot{\gamma}(z)  &=&  \frac{1}{ad} \sqrt{\phi g\cos\theta} (\tan\theta-\mu_0)   \left( (h-z)^{1/2}   - h_c\,  (h-z)^{-1/2}  ) \right) , \label{eq_gammadot}
\end{eqnarray}
while $ \dot{\gamma}(z) = 0$ for $z\geq h-h_c$. 

Thus, the velocity field $u(z)$ is computed by integration of equation~(\ref{eq_gammadot}), that leads for $z\leq h-h_c$ to:
\begin{eqnarray}
u(z) &=& u^{Bagnold}(z)+u^{cohesion}(z),  \nonumber 
 \end{eqnarray}
with 
\begin{eqnarray}
   \frac{u^{Bagnold}(z)}{\sqrt{gd}} &=&    \frac{1}{a} \frac{2}{3} \sqrt{\phi \cos\theta} \, (\tan\theta - \mu_0)\ \frac{h^{3/2} - (h-z)^{3/2}}{d^{3/2}}    \geq 0 
\end{eqnarray}
and 
 \begin{eqnarray}
 \frac{u^{cohesion}(z)}{\sqrt{gd}} &=& -  \frac{2}{a}  \sqrt{\phi \cos\theta}\ \frac{ h_c }{ d }\  \frac{h^{1/2}-(h-z)^{1/2}}{d^{1/2}} \leq 0. 
\end{eqnarray}
With equations~(\ref{eq_Imuaff2})-(\ref{eq_Imuaff3}), the velocity field can be alternatively written as:
 \begin{eqnarray}
  \frac{u(z)}{\sqrt{gd}} &=& \frac{2}{3} I^{Bagnold} \sqrt{\phi\cos\theta}\ \frac{h^{3/2} - (h-z)^{3/2}}{d^{3/2}}   \label{uIBagnold} 
  \end{eqnarray}
 \begin{eqnarray}
+2 I^{cohesion}(z) \sqrt{\phi\cos\theta}\ \frac{h-z}{d}\ \frac{h^{1/2} - (h-z)^{1/2}}{d^{1/2}}    \label{uIcohesion} 
\\  =  2 I^{Bagnold} \sqrt{\phi\cos\theta}\ \left(  \frac{1}{3} \frac{h^{3/2} - (h-z)^{3/2}}{d^{3/2}}  - \frac{h_c}{d} \ \frac{ h^{1/2} - (h-z)^{1/2}  }{d^{1/2}}  \right), 
\end{eqnarray}
 for $z\leq h-h_c$, while $u = u_c = u(h-h_c)$ for $z\geq h-h_c$:
 \begin{eqnarray}
    \frac{u_c}{\sqrt{gd} } &=& 
    \frac{2}{a} \sqrt{\phi \cos\theta} \ (\tan\theta - \mu_0) \ \left( \frac{1}{3}
     \frac{ h^{3/2}-h_c^{3/2} }{d^{3/2}} - \frac{h_c}{d} \  \frac{   h^{1/2} - h_c^{1/2}  }{d^{1/2}} \right) . 
    \end{eqnarray}
One example of a velocity profile and its two contributions are plotted in figure~\ref{fig_contributionBagCoh}b for 
$\tau_c=70$Pa, $\mu_0=0.45$, $a=0.7$, $\theta=33^{\circ}$ and $h=5$cm. 

The mass flow rate $Q_m$ follows: 
\begin{eqnarray}
 \frac{ Q_m} { Q_{m\,0} } & =&\frac{ h \overline{u}  }{ d\sqrt{gd} } \mbox{ with }  Q_{m\,0}=\phi\rho W d\sqrt{gd}.   \label{eq_qm1}
  \end{eqnarray}
 As for the velocity field, the mass flow–rate can be decomposed in different contributions: a Bagnold one and a cohesion one related to the velocity profile in the shear zone and a plug contribution in the plug zone: 
\begin{eqnarray}
 Q_m &=& Q_m^{Bagnold} + Q_m^{cohesion} + Q_m^{plug} . \label{eq_qm2}
 \end{eqnarray}
The continuity of the velocity at the interface $z=h-h_c$  leads to $Q_m^{cohesion}=-Q_m^{plug}$ so that $Q_m= Q_m^{Bagnold}$, simplifying the relation~(\ref{eq_qm2}): 
\begin{eqnarray}
  Q_m &=&   Q_{m\,0}\  \frac{1}{a} \frac{2}{3} \sqrt{\phi \cos\theta} \, (\tan\theta - \mu_0) \ \frac{ 3h^{5/2}/5 - h_c h^{3/2} + 2h_c^{5/2}/5 }{d^{5/2}},    \label{eq_qm3}
 \end{eqnarray}
with the following decompositions: 
 \begin{eqnarray}
(h-h_c) \, Q_m^{Bagnold} &=&  \frac{Q_{m\,0}}{ d\sqrt{gd}} \   h \int_0^{h-h_c} u^{Bagnold}(z) dz, 
\\ (h-h_c) \, Q_m^{cohesion} &=&   \frac{Q_{m\,0}}{ d\sqrt{gd}}  \   h \int_0^{h-h_c} u^{cohesion}(z) dz 
 \end{eqnarray}
and 
 \begin{eqnarray}
h_c\, Q_m^{plug} &=&   \frac{Q_{m\,0}}{ d\sqrt{gd}}  \    h \int_{h-h_c}^h u_c \,dz.
 \end{eqnarray} 
However, this is not a simple scaling between $Q_m$ and $h$, as it is for a dry granular flow, for which $h_c=0$: $Q_m^{dry}/ Q_{m\,0} \propto  \sqrt{\phi \cos\theta} \, (\tan\theta - \mu_0) (h/d)^{5/2} /  a $. 
At a constant $\theta$ (or $h$), the mass flow rate $Q_m$ increases for increasing $h$ (or $\theta$). 

\begin{figure}
	\centering
 	 a) \includegraphics[width=.45\linewidth]{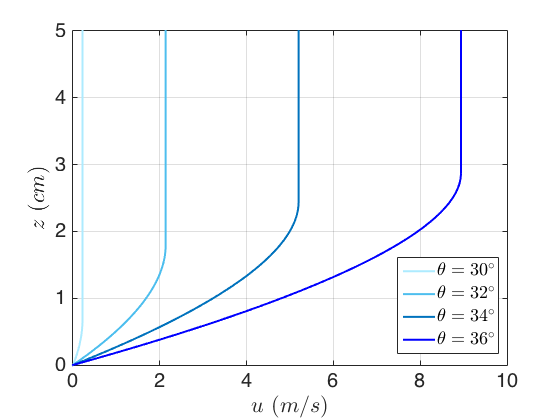} 
	 b) \includegraphics[width=.45\linewidth]{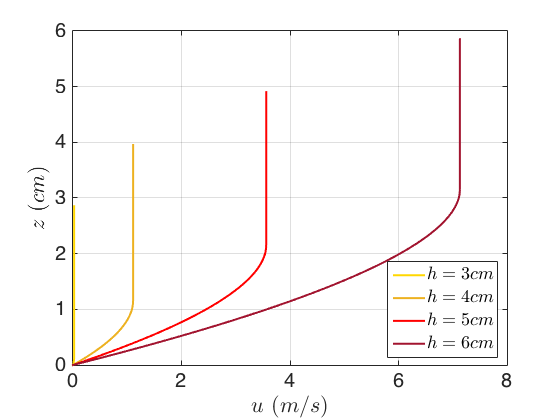} 
	\caption{Theoretical velocity profiles $u(z)$  of a wet granular flow for $\mu_0=0.45,\ a=0.7,\ \tau_c=70$Pa: two series at a constant height $h=5$cm (a) and a constant slope angle $\theta=33^{\circ}$ (b).  }
	\label{fig_uztheo}
\end{figure}

Figure~\ref{fig_uztheo} shows some theoretical profiles $u(z)$ for two series  of ($h$, $\theta$) --at constant height $h=5$cm and constant slope angles $\theta=33^{\circ}$--, computed from equations~(\ref{uIBagnold})-(\ref{uIcohesion}), typical of a shear flow topped with a plug flow. 

Figure~\ref{fig_contributionBagCoh} shows that both Bagnold and cohesion contributions to the wet granular flow inertial number and velocity profiles 
are of the same order (but of opposite signs), so that none is negligible. 
As an example,  for the set of parameters ($\theta=33^{\circ}$, $h=5$cm, $\tau_c=70$Pa, $\mu_0=0.45$ and $a=0.7$), $-I^{cohesion}$ is between $0.16$ and $0.28$ for $z\in[0$-$h-h_c]$ compared to $I^{Bagnold}=0.28$ (figure~\ref{fig_contributionBagCoh}a). This is still true for all of our experimental parameters. This leads to a velocity profile $u(z)\geq0$ that can not be approached by one or the other contribution $u^{Bagnold}(z)\geq0$ or $u^{cohesion}(z)\leq0$ (figure~\ref{fig_contributionBagCoh}b), 
 preventing us from deducing simple scaling laws relating $I(z)$, $u(z)$ and $Q_m$  from their analytic relations. However, computations done just below, will bring us some knowledge.

\begin{figure}
	\centering
 	 a) \includegraphics[width=.45\linewidth]{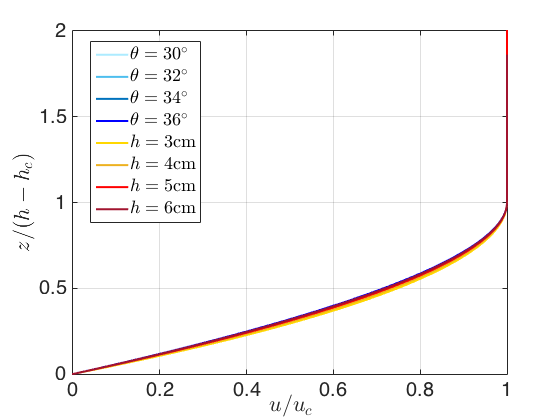} 
	b) \includegraphics[width=.45\linewidth]{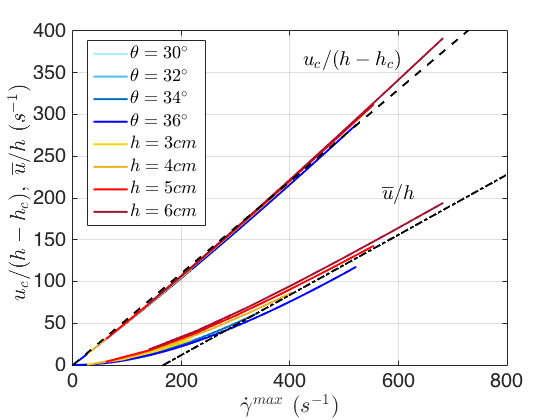} 
	\caption{ a) Several velocity profiles $u(z)$ of a wet granular flow normalized by the plug velocity $u_c=u(h-h_c)$ and the shear zone length $(h-h_c)$ at  $h=5$cm and $\theta=33^{\circ}$. b)  Two estimations for a characteristic shear rate: $u_c/(h-h_c)$ and $\overline{u}/h$ as a function of the local shear rate at the bottom    $\dot{\gamma}^{max}=\dot{\gamma}(z=0)$. }
	\label{fig_utheo}
\end{figure}

\begin{figure}
	\centering
	a) \includegraphics[width=.45\linewidth]{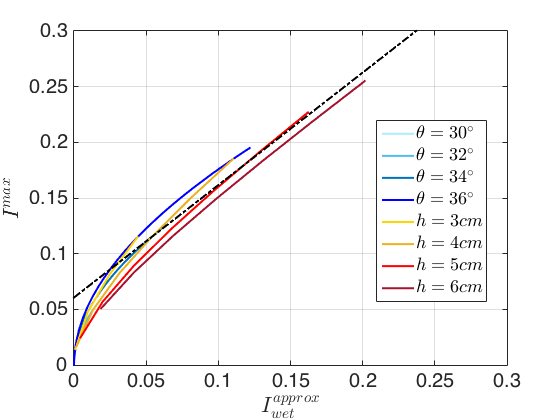} 
	b) \includegraphics[width=.45\linewidth]{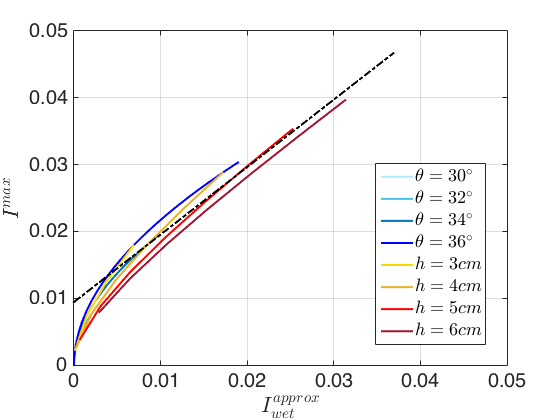} 
	\caption{ Inertial number $I^{max}=I(z=0)$ (local value at the bottom)  as a function of the approximation
	$I^{approx}_{wet}$ computed from equation~(\ref{eq_Iwet}) for $\mu_0=0.45$  
	 and $\tau_c=100$Pa for series at constant $\theta=33^{\circ}$ ($h=3,\ 4,\ 5$ and $6$cm) and constant $h=5$cm ($\theta=30,\ 32,\ 34$ and $36^{\circ}$) but for two different values of $a$: a) $a=0.7$ and b) $a=4.5$. All the plotted data are proportional to $1/a$. 	}
	\label{fig_I}
\end{figure}

\begin{figure}
	\centering
	\includegraphics[width=.45\linewidth]{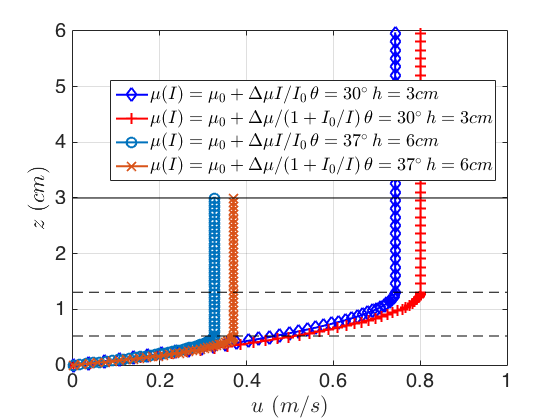}
	\caption{ 	Two examples of normal profiles $I(z)$ and $u(z)$ assuming either an affine function ($a=\Delta\mu/I_0$) or a more elaborate model for $\mu(I)$ with the parameters $I_0=0.38$, $\mu_0=0.41$ and $\Delta\mu=0.35$ from~\citep{Saingier16}
	 and $\tau_c=100Pa$ for ($\theta=30^{\circ},\ h=6$cm) and ($\theta=37^{\circ},\ h=3$cm).
	}
	\label{fig_Icomp}
\end{figure}

When using the characteristic velocity $u_c$ at the interface and its position $h-h_c$ to non-dimensionalize computed velocity profiles $u(z)$, data of  $u(z)/u_c$  as a function of $z/(h-h_c)$ on the shear zone ($z\leq h-h_c$)
 approximately collapse  for different $h$ and $\theta$  at a given rheology (figure~\ref{fig_utheo}a). 
Moreover, we observe a proportionality between the $z$-averaged velocity $\overline{u}$  
and the plug velocity $u_c$ for several sets of $h$ and $\theta$, typical of our experiments: $u_0\simeq0.8u_c$ with a  very good correlation coefficient $R=0.99$ (figure not shown here).   

In addition, from our experimental measurements made on a wet granular flow ($h$, $u_0$ and $Q_m$), one would like to compute one characteristic inertial number $I$ (e.g. $I^{max}=I(z=0)$) even without the knowledge of the parameters $\mu_0$ and $a$ of the $\mu(I)$ function, as it can be done for a dry granular flow thanks to the relations: 
\begin{equation}
I_{dry} = \frac{5}{2}  \frac{1}{\sqrt{\phi\cos\theta}} \frac{Q_m}{Q_{m\,0}} \frac{d^{5/2}}{h^{5/2}}  = \frac{5}{2}  \frac{1}{\sqrt{\phi\cos\theta}} \frac{\overline{u}}{\sqrt{gd}} \frac{d^{3/2}}{h^{3/2}} .  \label{eq_Idry}
\end{equation}
However, for unsaturated granular material, one needs a priori the knowledge of $\tau_c$, $\mu_0$ and $a$ to compute $I_{wet}$, as shown by relations~(\ref{eq_Imuaff})-(\ref{eq_Imuaff3}) or (\ref{eq_qm3}). 
To overpass this difficulty, we try to estimate the shear rate at the bottom $\dot{\gamma}(z=0)=\dot{\gamma}^{max}$ with measurements that can be achieved in our experiments, in order to estimate $I^{max}$. We observe in figure~\ref{fig_utheo}b a non expected linear relation between $\dot{\gamma}^{max}$ and $u_c/(h-h_c)$, that are respectively the local shear rate at the bottom and some 'averaged' shear rate in the shear layer (approximated as the one for a linear velocity profile between $0$ and $u_c$): $u_c/(h-h_c)\simeq0.55\, \dot{\gamma}^{max}$ with a  very good correlation coefficient $R=0.99$. 
But still, we have not access in our experiments to the measurement $u_c/(h-h_c)$. 
More interestingly, there is approximately an affine relation between  $\dot{\gamma}^{max}$ and $u_0/h$ --that is  achievable in our experiments--  for large values ($\dot{\gamma}^{max}\geq300$s$^{-1}$ and $u_0/h\geq50$s$^{-1}$), as illustrated in  figure~\ref{fig_utheo}b where $u_0/h\simeq 0.36\, \dot{\gamma}^{max}-60$. 
Note that the proportionality coefficient $1/0.36=2.78$ between $\dot{\gamma}^{max}$ and $u_0/h$ is not far away from $2.5$, the coefficient in the case of a pure Bagnold profile, as seen in equation~(\ref{eq_Idry}).
This approximation (even if not perfect) can be used to compute the characteristic inertial number $I^{max}$ in our experiments  as:
\begin{eqnarray}
I^{approx}_{wet} &\simeq& 2.78 \frac{1}{\sqrt{\phi\cos\theta}} \frac{Q_m}{Q_{m\,0}} \frac{d^{5/2}}{h^{5/2}}  = 2.78  \frac{1}{\sqrt{\phi\cos\theta}} \frac{\overline{u}}{\sqrt{gd}} \frac{d^{3/2}}{h^{3/2}} .  \label{eq_Iwet} 
\end{eqnarray}
Using equation~(\ref{eq_qm3}), this gives the relation:
\begin{equation}
I^{approx}_{wet} \simeq  2.78 \  \frac{1}{a} \frac{2}{3}  \, (\tan\theta - \mu_0) \  \left( \frac{3}{5} - \frac{h_c}{h} + \frac{2}{5} \left(\frac{h_c}{h} \right)^{5/2} \right)  
\end{equation}
as an approximation of equation~(\ref{eq_Imax}). 
We check in figure~\ref{fig_I}a  for our theoretical data that this approximation is well correlated with  the characteristic inertial number $I^{max}$, 
 even if neither perfectly equal nor perfectly proportional: the approximation underestimates slightly the characteristic inertial number of the wet granular flow, as $I^{max}\simeq I^{approx}_{wet}+0.07$. However, one can extrapolate that for high values of $I$ (meaning $I>0.07$  for $\mu_0=0.45$ and $a=0.7$), both are equal: $I^{max}\simeq I^{approx}_{wet}$.  
By default on doing better, we will use the approximation~(\ref{eq_Iwet}) for our experimental wet granular flows. Note that a better precision can be reached by using $I^{approx}_{wet}+0.07$ if $\tau_c=70$Pa, $\mu_0=0.45$ and $a=0.7$.

Finally, in order to generalize the presented computations, one can notice that the inertial number $I$, the velocity $u$ and the mass flow rate $Q_m$ are proportional to $1/a$, so that all the results reported here for one value of $a$ can be used to deduce results for a different value of $a$. Increasing $a$ by a factor $k$ at constant $\mu_0$, leads to decrease $I$ and $u$ by a factor $1/k$. 
In particular, this allows to generalize the approximation $I^{max}\simeq I^{approx}_{wet}$ in the case of $\mu_0=0.45$ for all values of $a$, at the condition that $I^{max}>0.07*0.7/a$. As an example, 
 figure~\ref{fig_I}b shows $I^{max}(I^{approx}_{wet})$ for $\tau_c=70$Pa, $\mu_0=0.45$ and $a=4.5$, checking that the approximation $I^{max}\simeq I^{approx}_{wet}$ for typical values of $I>0.01$. 

\subsection{A more elaborate function $\mu(I)= \mu_0 +\frac{\Delta\mu}{1+\frac{I_0}{I}}   $ }

All the previous computations performed for the affine function of $\mu(I)$  can be done for a more elaborate model with three parameters as commonly used in the literature~\citep{Jop06}. 
Such a function allows to describe the saturation of the internal friction with large inertial numbers, contrary to the affine function, leading to some accelerating flows. In this case, the inertial number follows as:  
\begin{eqnarray*}
 I^{3\,parameters}(z) &=& I_0 \frac{\mu-\mu_0}{\Delta\mu+\mu_0-\mu} 
= I_0 \frac{h-h_c-z}{(h-z)(1/\delta-1)+h_c} 
\end{eqnarray*}
where $\delta=(\tan\theta-\mu_0)/\Delta\mu$, leading to the local shear rate as: 
\begin{eqnarray*}
 \dot{\gamma}(z) &=&  \frac{I_0}{d} \sqrt{\phi g\cos\theta} \frac{ (h-z)^{1/2}(h-h_c-z) }{ (h-z)(1/\delta-1) +h_c }
\end{eqnarray*}
 and the velocity field as: 
\begin{eqnarray*}
 u(z) &=&  k_1 (h - z)^{3/2} + k_2 (h - z)^{1/2} + k_3\ \arctan(k_4 (h - z)^{1/2}) + k_5 , 
\end{eqnarray*}
with five constants $k_i$ (with $i=1,2,3,4$ and $5$) dependent on the  parameters of the rheology and of the flow. 
However, the expressions are not so compact and easy to manipulate as previously, e.g. a term $k_3\arctan(k_4(h-z)^{1/2})$ appears in the expression of $u(z)$.  
Most importantly, when we compare quantitatively theoretical predictions from these two laws, there are only few non-significant differences in the range of our experimental values (figure~\ref{fig_Icomp}). 

If we define $a$ in the affine function as $a=\Delta\mu/I_0$, with $\Delta\mu$ and $I_0$ from the more elaborate model, as suggested by an asymptotic development, then one systematically gets $$I^{affine} = I^{3\,parameters} \left(  1-\delta +\delta\ \frac{h_c}{h-z} \right) \leq I^{3\,parameters}, $$
that would lead to predictions from  the  elaborate model for velocities and mass-flow rates, slightly larger than from the affine model. 

After having derived and computed analytical solutions of a steady uniform free surface gravitational cohesive frictional flow, whose rheology is encoded by an $I$-extended Mohr-Coulomb plastic law in equation~(\ref{eq_rheocohesion}), we will analyze our experimental results within this theoretical framework. 

\begin{figure}
	\centering
	 a) \includegraphics[width=.45\linewidth]{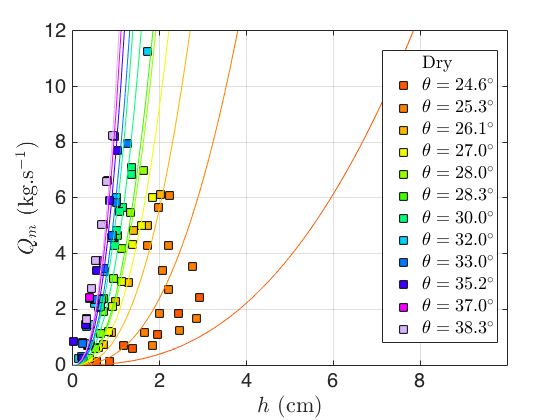}	 	 
	 b) \includegraphics[width=.45\linewidth]{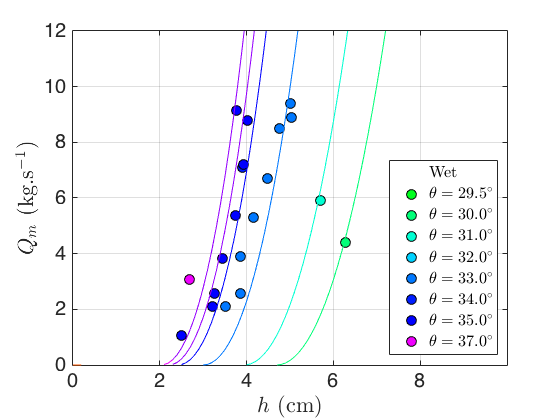}	 	 
\caption{Dry and wet granular flows. Mass flow rate $Q_m$ experimentally measured as a function of the thickness of the dry (a) and wet (b)  steady uniform granular flow $h$ for the experiments reported in this paper. Continuous  lines are predictions from equation~(\ref{eq_qm3}) with $\mu_0=0.46$, $\tau_c=70$Pa, $a_{dry}=0.7$ and $a_{wet}=4.5$. Colors encode for the incline slope angle $\theta$ as in figure~\ref{fig_param}. }
	\label{fig_Qh}
\end{figure}

\section{Experimental results}
\label{results}

\subsection{Steady and uniform wet or dry granular flows over a rough incline}

We choose values of our control parameters  $f$, the aperture thickness and $\theta$,  the slope angle, so that the resulting flow far behind the front is steady and uniform, at constant height $h$ and mean velocity $\overline{u}=u_0$, the front velocity. Indeed such flows are not observed for too small values of $f$ (or  $h$) and $\theta$ and these minima for wet granular flows are  larger than the ones for dry granular flows (figure~\ref{fig_param}). 
The steadiness of the flow can be seen  from the constant mass flow rate measured out of the incline (figure~\ref{fig_massflowrate}a) and from the constant velocity $u_0$ of the flow front (figure~\ref{fig_hmeas}b).  

At a constant slope angle $\theta$, the steady uniform height  $h$ and the steady velocity of the flow $u_0$ increase with the aperture height $f$. At a constant aperture height $f$, $h$ decreases with $\theta$ while $u_0$ increases. As a result, the mass flow rate  $Q_m$ increases with the slope angle $\theta$ (or the aperture thickness $f$) at constant $f$  (or at constant $\theta$), as shown in figure~\ref{fig_Qh}a and b. 
The same evolution is observed for dry and wet granular flows. 
The experiments reported here have a mass flow rate $Q_m$ of the same order of magnitude  whether  wet or dry. 

Furthermore, we check  that our measurements of $h$ and $u_0$ are coherent with the independent measurement of the mass flow rate $Q_m \simeq \phi\rho W h u_0$, in which $W$ is the width of the incline. In the following, we will prefer to use the measured $Q_m$ (associated to $h$) instead of $u_0$ when possible, because this former is less sensitive to experimental noise because of its  integration in time and space. 

\begin{figure}
	\centering
	 a) \includegraphics[width=.45\linewidth]{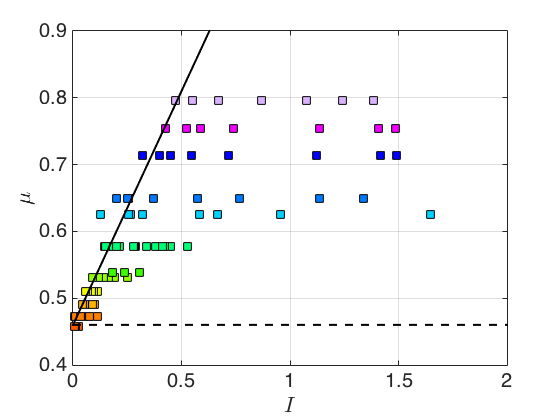}	 	 
	 b) \includegraphics[width=.45\linewidth]{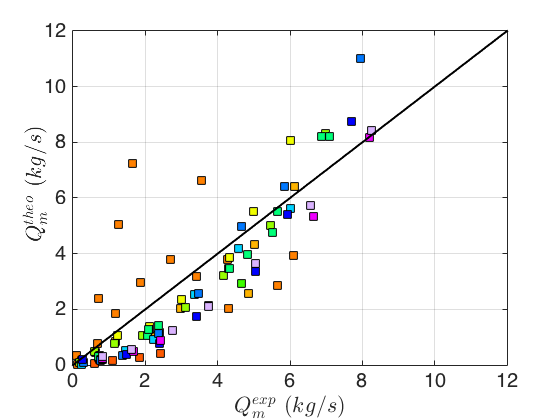}
\caption{Dry granular flows. a) Friction coefficient $\mu=\mu^{eff}=\tan\theta$ as a function of the inertial number $I$ (equation~(\ref{eq_Idry})). b) Theoretical values of the mass flow rate $Q_m$ as a function of the experimental ones for the rheological parameters $\mu_0=0.46$ and $a_{dry}=0.70$ optimizing their linear correlation. Both bold lines in (a) and (b) are for $\mu_0=0.46$ and $a_{dry}=0.70$. Colors encode for the incline slope angle $\theta$ as in figure~\ref{fig_param}. }
	\label{fig_dry}
\end{figure}

\subsection{Identification of the rheological parameters for dry granular flows}
\label{subsec_muidry}

Our experiments of steady and uniform dry granular flows over a rough inclined plane behave as expected from the literature~\citep{Pouliquen99,GDRMiDi04}. All of the theoretical relations written above for cohesive frictional materials apply to dry (cohesionless) flows by taking the stress threshold $\tau_c=0$. In particular, equation~(\ref{eq_Idry}) gives the inertial number $I$ as a function of measurements experimentally achievable, allowing to plot the friction coefficient  
as a function of $I$ in figure~\ref{fig_dry}a.  As expected, the friction coefficient $\mu$ of the dry granular material increases with the inertial number $I$. 
 As in~\citep{Pouliquen99,GDRMiDi04,Saingier16}, experimental $(\mu$, $I)$ data show some scattering, that was recently attributed to non-local effects~\citep{Perrin21}. 
 Thus a  fit by an affine function  gives rather bad precision on the rheological parameters. In particular, such direct fit  is very sensitive to the ranges of slopes and heights $(\theta$, $h)$ experimentally explored. 
Whereas the value of  $\mu_0$ is rather confident:  $\mu_0\simeq0.46\pm0.01$, the slope $a$ is not: $a\simeq 0.5 \pm0.2 $. Still, this gives the order of magnitude for $a$.  

Alternatively, we define our fit parameters ($\mu_0$, $a$) as the ones optimizing the correlation between the theoretical --from equation~(\ref{eq_qm3}) with $h_c=0$-- and experimental values of the mass flow-rate $Q_m$.  We found that the affine function characterizing the best our dry granular flows has 
 $\mu_0=0.46$  and $a=0.7$. For these parameters, the theoretical 
  and experimental values of the mass flow-rate are plotted in figure~\ref{fig_dry}b, as well as $\mu(I)$ in figure~\ref{fig_dry}a and theoretical curves $Q_m(h)$ computed  from equation~\ref{eq_qm3} for slope angles explored here in figure~\ref{fig_Qh}a. 
 This optimization of the affine parameters is robust in the sense that the linear relation between theoretical and experimental $Q_m$ has its scattering (quantified by the precision of a linear fit) directly related to the value of  $\mu_0$ only, while its slope is directly related to $a$ only, as can be deduced from equations~(\ref{eq_hc}) and~(\ref{eq_qm3}). 
We checked that the variation of the solid fraction $\phi$ with the inertial number $I$ (according to $\phi(I)=\phi_{RCP}-\Delta\phi\, I$ with $\Delta\phi$=0.1 and $\phi_{RCP}\simeq0.6$ from~\citep{Fall15}) to compute the theoretical mass flow-rate, does not change our results. 

In the following, we will test whether or not this rheological law $\mu(I)=0.46+ 0.7 \, I$, with parameters identified from our dry granular flows will allow us to predict quantitatively wet granular one. The proportional coefficient between $\mu$ and $I$ will be referred to as $a_{dry}=0.7$. To this aim, we will need to estimate the cohesion stress $\tau_c$ first. 
 
\begin{figure}
	\centering
	 a) \includegraphics[width=.45\linewidth]{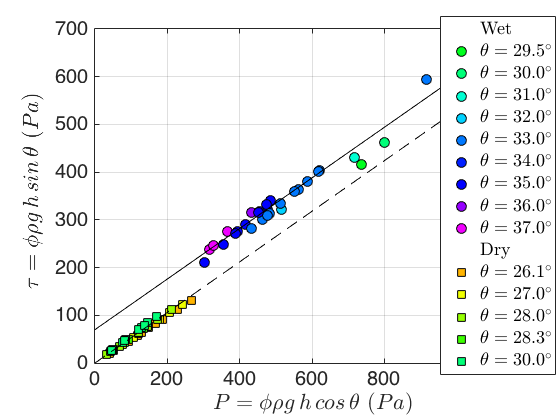} 	  
 	 b) \includegraphics[width=.45\linewidth]{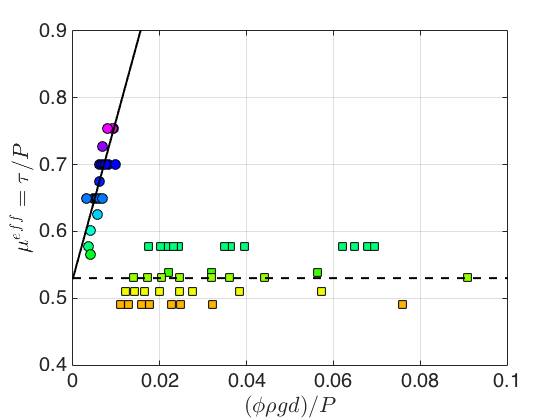} 
	\caption{Wet and some dry granular flows.  a) Shear and normal stresses $\tau$ and $P$ for all of  our wet flows and for dry flows at $\theta\in[26$-$30]^{\circ}$, equivalent to $\mu\in[0.5$-$0.6]$, such that they have approximately  both the same internal friction $\mu$. 
	b) The same data are shown in a different representation: the effective friction $\tau/P$ as a function of the normalized inverse pressure $\phi\rho g d/P$. Colors encode for the incline slope angle $\theta$ as in figure~\ref{fig_param}. }
	\label{fig_tauPselect}
\end{figure}

\begin{figure}
	\centering
 	 a) \includegraphics[width=.45\linewidth]{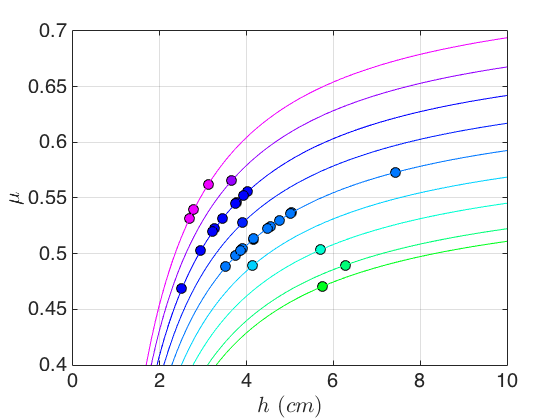} 	 
  	 b) \includegraphics[width=.45\linewidth]{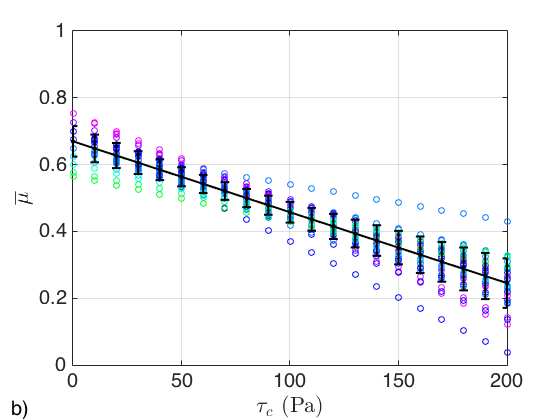} 	  
	\caption{Wet granular flows. a) Theoretical values of the internal friction $\mu(z=0)$ as a function of height $h$ for our experiments assuming  	$\tau_c=70$Pa from equation~(\ref{eq_muz}). Such a computation is systematically done for various  $\tau_c$ to construct figure~b. 
	b) 	Mean value $\overline{\mu}$, with the standard deviation encoded in the errorbar, of the internal friction $\mu$ for our experiments (as well as the single values $\mu$  of all of our realizations in colored points) as a function of the value assumed for $\tau_c$. Colors encode for the incline slope angle $\theta$ as in figure~\ref{fig_param}.  	}
	\label{fig_muwet}
\end{figure}

\subsection{Wet granular flows: cohesion and friction } \label{wetrheo}
\subsubsection{Relevance and measurement of the cohesion stress $\tau_c$}

In order to characterize the cohesion stress  $\tau_c$, 
we drawn in figure~\ref{fig_tauPselect}a the coordinates $(P,$ $\tau)$ --the pressure and the shear stress at the bottom $z=0$ from equations~(\ref{eq_P}) and~(\ref{eq_tau})-- for the wet material (circles). All of our wet experiments lie along a single line, allowing to state that  they all have  approximately the same internal friction $\mu\simeq0.53$. Moreover they highlight a well defined cohesion stress $\tau_c\simeq70$Pa.  
For comparison, we plot the coordinates $(P,$ $\tau)$ for the dry material (squares) when friction is about $\mu\simeq0.53$, that is for slopes $\theta\in[26$-$30]^{\circ}$. 
As expected from figure~\ref{fig_param} showing steady uniform dry flowing layers  thinner than wet ones,  $(P,$  $\tau)$ data of the dry glass beads lie below  the wet ones: there is a slight shift between the two curves indicating that the effective friction coefficient of the wet beads is higher than 
the dry ones. 
Moreover dry data pass through the origin indicating no cohesion within the bulk. 

Notice that each point (both for dry and wet flows) is obviously computed from equations~(\ref{eq_P}) and~(\ref{eq_tau}) for $z=0$. 
Thus, each single point has its ratio $\tau/P=\tan\theta$, and each set of data realized at a constant slope angle $\theta$ is aligned according a straight line of slope $\tan\theta$  passing through the origin. Also, note that even if we do not consider $\phi$ as a constant but dependent on $I$, we checked that this does not change our results. 
For example, green points in figure~\ref{fig_tauPselect}a correspond to wet and dry flows at $\theta=30^{\circ}$, so they are aligned along the straight line passing through the origin and of slope $\tan\theta=\tan30^{\circ}\simeq0.58$. Surprisingly, our experimental wet data demonstrate that the material can be straightforwardly described by an affine law, indicating  a finite cohesion stress $\tau_c$ and a constant internal friction $\mu$. 
So, for dry granular flows, we had to select a small range of slopes $\theta\in[26$-$30]^{\circ}$ in order to get the same internal friction. 

We now probe of what allows to state that 
$(P,$ $\tau)$ wet data 
have the same internal friction  $\mu$. 
Two experiments at controlled parameters $(\theta,$ $h)$ 
have the same internal friction $\mu$, not if they are at the same $\theta$ (as for dry flows), but if 
\begin{eqnarray}
\mu =\frac{\tau-\tau_c}{P}=  \tan\theta - \frac{\tau_c}{\phi\rho g h\cos\theta} \label{eq_muz2} 
\end{eqnarray} is constant.  
Notice that, this depends only on the single rheological parameter $\tau_c$.
Thus, assuming that equation~(\ref{eq_rheocohesion}) holds,  
figure~\ref{fig_muwet}a shows the value of $\mu$ as a function of $h$ for a given $\tau_c=70$Pa  
and for different constant values of $\theta$. 
One can see that $\mu$ should be between $[0.47$-$0.57]$ with a mean value of $0.52$ and a standard deviation equal to $\pm0.03$.   
This computation is systematically done for various values of $\tau_c\in[0$-$200]$Pa; the mean values, standard deviations and individual values of $\mu$ for our explored experimental parameters are shown in figure~\ref{fig_muwet}b. 
For example, for $\tau_c=0$Pa, equivalent to a dry granular flow, one retrieves that $\mu=\tan\theta$; for finite values of $\tau_c\neq0$, equation~(\ref{eq_muz2}) shows that $\mu$ decreases with $\tau_c$, and increases with $h$ (with $\theta$ respectively) at a constant $\theta$ ($h$ respectively). 
 Figure~\ref{fig_muwet}b shows that  our wet experiments are approximately at the same internal friction whatever the value of $\tau_c$ would be, 
 with a typical standard deviation of $\pm0.03$. 
Finally, figure~\ref{fig_muwet}b demonstrates that  for $\tau_c=70$Pa, internal frictions $\mu$ of our wet experiments have the smallest  standard deviation. 

Another way of representing these wet and dry data  
is to plot the effective friction $\mu^{eff}=\tau/P$ as a function of the dimensionless inverse pressure $\phi\rho gd/P$, that is nearly constant for dry flows but shows a linear relation that may be used to identify the value of cohesion stress $\tau_c=70$Pa. The straight lines intercept the vertical axis at nearly the same values for dry and wet flows $\mu\simeq0.53$: for large pressures, cohesive capillary forces become negligible, and the material behaviour should be the same as in the absence of the wetting fluid~\citep{Badetti18,Badetti18b}. 

As a summary, we demonstrate that for one liquid content $\epsilon$, our wet granular flows are well described by a relation $\tau=\tau_c+\mu P$, consistently with our dry granular flows described by a relation $\tau=\mu P$, by carefully considering dry and wet flows at the same value of $\mu$. This allows to measure the cohesion stress as $\tau_c\simeq70$Pa. 
At this step, we do not know yet if the same function $\mu(I)$ describes both dry and wet data. 

\begin{figure}
	\centering
	 a) \includegraphics[width=.45\linewidth]{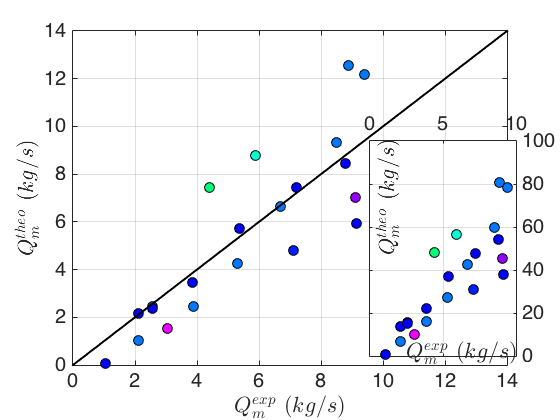}
	 b) \includegraphics[width=.45\linewidth]{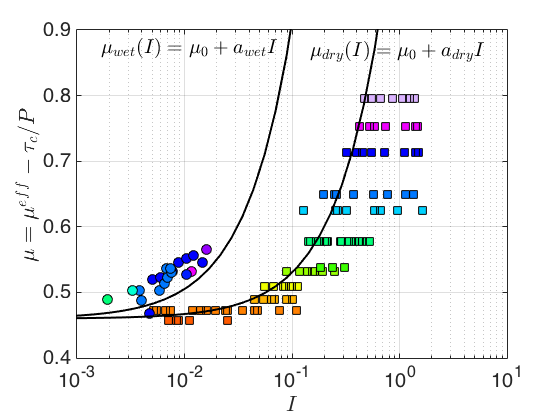}
\caption{a) Theoretical and experimental values of the mass flow rate $Q_m$ for our wet flows for two sets of rheological parameters for the internal friction  $\mu(I)=\mu_0+a\,I$: $a=4.5$ and $\mu_0=0.46$ in the main panel are obtained from the optimization of the linear correlation, whereas $a=a_{dry}=0.7$ and $\mu_0=0.46$ used  in inset come from our dry flows. 
b) Experimental data from dry and wet granular flows: internal friction $\mu(I)$ as a function of the characteristic inertial number $I$. Colors encode for the incline slope angle $\theta$ as in figure~\ref{fig_param}. }
	\label{fig_wet}
\end{figure}

\subsubsection{Internal friction relation $\mu(I)$}

If the Mohr-Coulomb criterion as in~\citep{Pierrat98,Badetti18,Badetti18b} applies, then $\mu(I)$ 
coincides with the internal friction coefficient of the dry material already identified in section~\ref{subsec_muidry} as $\mu(I)=0.46+0.7\,I$. 

We turn now to the wet material, and compute the mass flow rate from equation~(\ref{eq_qm3}) with these parameters for $\mu(I)$ and for $\tau_c=70$Pa and compare it with our experimental measurements. In insert of figure~\ref{fig_wet}a, we can see a very good correlation between theoretical and experimental data but with a slope of $\approx10$. 
These theoretical values over-estimate significantly the mass flow rate, as well as the front velocity, the shear rate and the inertial number of our wet flows, when using the frictional law of our dry sample. 

By using  the same method as  
in section~\ref{subsec_muidry}, we identify the best parameters of $\mu(I)$ from the optimization of the linear correlation between theoretical and experimental values of mass flow rates. This is also possible here for our wet flows, because again, flow properties are proportional to $1/a$, so that the goodness of fit $Q_m^{theo}(Q_m^{exp})$ depends on $\mu_0$ only, while the subsequent linear correlation with a constant equal to $1$ depends on $a$ only. This method has the great advantage of being independent on the definition chosen for $I$, so that the parameters identified here are robust with respect to any approximation and depend only on the whole set of ($Q_m,h,\theta$) data.  We found that the best affine law of $\mu(I)$ for our wet flows is:  $\mu(I)=0.46+4.5\,I$, as shown in figure~\ref{fig_wet}a. For these parameters, the theoretical curves $Q_m(h)$ are computed from equation~(\ref{eq_qm3}) for slope angles explored here in figure~\ref{fig_Qh}b. 

This result can be inferred too from figure~\ref{fig_wet}b where 
the internal friction $\mu$ is plotted as a function of $I$ for dry and wet flows. 
We clearly see that the function $\mu(I)$ for wet grains is larger and increases much more quickly with $I$ than for dry flows.


\section{Discussion}
\label{disc}

First, we come back to the choices we made for the internal friction dependence with the inertial number $\mu(I)$ presented in section~\ref{theo}.

\subsection{Constant value $\mu(I)=\mu_0$ for wet flows: a too crude approximation}

When computing the maximal value of $\mu$  
as $\mu^{max}=\mu(z=0)$ from equation~(\ref{eq_muz}) or~(\ref{eq_muz2}), 
we observe that $\mu(I)\simeq\mu_0$  in our wet experiments (figures~\ref{fig_tauPselect} and \ref{fig_wet}b). 
However such an approximation in the framework of the constitutive law~(\ref{eq_rheocohesion}) --with friction and cohesion-- would lead to huge relative errors on stresses  
larger than $100\%$.  
Indeed, the pressure and shear stress in the wet granular layer would be approximated as constant: $P(z) \simeq P_0 = \tau_c/(\tan\theta-\mu_0)$ and $\tau(z) \simeq \tau_0 = \tau_c\, \tan\theta/(\tan\theta-\mu_0) $. 

By contrast, this approximation would be not so crude 
in the case of a dry flow, for which the linear dependence of stresses with $z$  is recovered: 
 $\tau(z)= \tan\theta\, P(z)\simeq\tan\theta_0\, P(z) $ and $P(z)\simeq P(z) \cos(\theta_0)/\cos(\theta)$,  leading  to smaller relative errors ($\leq 40\%$ for $\theta\in[30$-$40]^{\circ}$) as much as $\Delta\theta/\theta_0\approx40\%\ll1$ with $\Delta\theta$ the range of explored slope angles. 
In this context, it is not relevant to compute predictions for our wet flows by considering $\mu(I)\simeq\mu_0$. 

Whenever we would consider $\mu(I)\simeq\mu_0$, computing the shear rate (and the inertial number $I$) would need an additional constitutive relation.

\subsection{A modified Bagnold law}

A possible and common candidate for dry granular flows would be the Bagnold law~\citep{Silbert01,Staron10}. 
\citet{Brewster05} suggested a modified Bagnold law for cohesive flows: 
$\tau =  \kappa \dot{\gamma}^2  + \eta \dot{\gamma}  $, allowing to compute velocity profiles. 
However, comparisons with their cohesive discrete numerical simulations show that 
  $\kappa$ and $\eta$ are not constant, but  are computed as fit parameters in the absence of theory about their evolution. An 
  alternative would have been to  modify the Bagnold law for cohesive flows as:  
$\tau =  \tau_c + \kappa \dot{\gamma}^2  $. 
Given the current knowledge, we choose to focus on the framework of the $\mu(I)$ rheology instead of the Bagnold law. 

Then, we will comment the distinct values $a_{dry}$ and $a_{wet}$  
found to describe  the internal friction $\mu$ as a function of the inertial number $I$.

\subsection{Liquid bridges induce cohesion, but enhance internal friction too}

One of our main major result is that cohesion induced by liquid bridges in a granular material (in the pendular state, with the liquid content being $\epsilon=0.5\%$) is well accounted for by equation~(\ref{eq_rheocohesion}), that can be seen as an extension of the classical Mohr-Coulomb plastic law with the internal friction dependent on the inertial number $I$. 
The idea behind this rheology is to distinguish the cohesion of the material encoded in the stress threshold $\tau_c$ and the dissipation encoded in the friction $\mu$. 
By comparing quantitatively the internal friction of dry and wet flows in the same configuration of an incline rheometer and using the same beads. We have shown however that dry and wet samples have straightforwardly different internal friction laws $\mu(I)$. 
This means that, under flow, the presence of liquid bridges not only induces cohesion, but modify  
the internal friction. 

This modification of the internal friction may not be due to the modification of the friction at the grain scale,  but 
to the modification of geometrical microstructural informations  (fabric tensor, coordination number, ...) and can be related to the nature of liquid bridges~\citep{Gans20}, that may change friction properties~\citep{Scheel08,Badetti18}.

Now, we will discuss about some scaling that should allow for a collapse of the two quantitatively distinct rheological curves $\mu(I)$ for wet and dry samples. 

\subsection{The unified visco-cohesive inertial number $I_m$ from~\citet{Vo20} } 

As seen in figure~\ref{fig_wet}b, the internal friction coefficient $\mu$ appears to grow faster with $I$ for  wet flows than  dry ones.One could wonder if this is due to the use of the inertial number $I$ defined in equation~(\ref{I}) as the  relevant one  for  cohesive materials, that would introduce some bias: is it possible that it under-estimates the relevant dimensionless number preventing a collapse of data for our wet and dry flows?  
Let us recall that $I$ allows to compare contributions of inertial and relaxation effects, that are here due to shear and confinement respectively, so that $I$ can be interpreted as the ratio of the related time scales $t_r=d / \sqrt{P/\rho}$ and $t_i=\dot{\gamma}^{-1}$, or the ratio of the related stress scales $\sigma_r=P$ and $\sigma_i=\rho d^2 \dot{\gamma}^2$: 
 $I=t_r/t_i=\sqrt{\sigma_i/\sigma_r}$. The presence of cohesion may induce some additional time and shear scale. 
This is what is done by~\citet{Vo20}, who change the relaxation stress by a linear combination of the pressure and  the elementary cohesion stress (at the particle scale) to take into account cohesion. 
To be much more general, they take into account viscous effects of a surrounding fluid too. 

\begin{figure}
	\centering
	a)  \includegraphics[width=.45\linewidth]{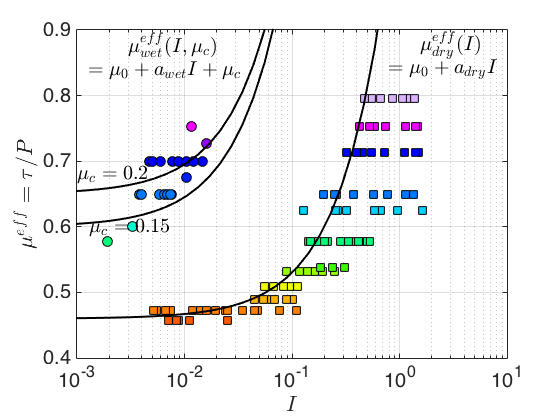} 
	 b) \includegraphics[width=.45\linewidth]{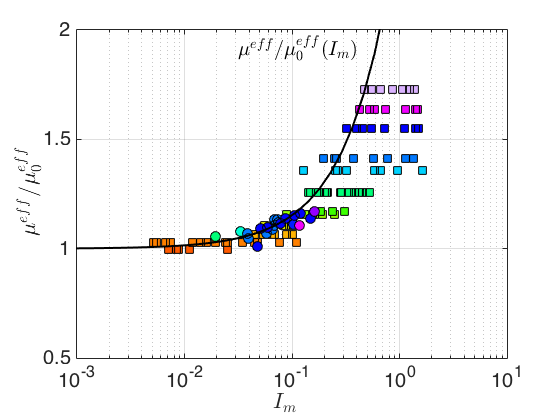} 
\caption{Wet and dry granular flows. a) Effective friction $\mu^{eff}=\tau/P$  as a function of the characteristic inertial number $I$. b) Normalized effective friction $\mu^{eff}/\mu^{eff}(I=0)$  as a function of the modified inertial number $I_m$ defined by~\citet{Vo20}. Colors encode for the incline slope angle $\theta$ as in figure~\ref{fig_param}. 
}
	\label{fig_mueffIwetdry}
\end{figure}

The authors defined an extended unified visco-cohesive inertial number $I_m$: 
\begin{eqnarray}
 I_m &=& I \left(\frac{1+\beta/St}{1+\alpha\xi}\right)^{1/2},   \label{eqImgal}
\end{eqnarray}
where $St$ is the Stokes number and $\xi$ is the cohesion index, $\alpha$ and $\beta$ being up to now unknown material-dependent parameters describing the additivity of frictional, cohesion and viscous effects to the rheology. The approach adopted by~\citet{Vo20} consist of expressing the effective friction coefficients normalized by their quasi-static values, denoted by $\widetilde{ \mu^{eff}}= \mu^{eff} /  \mu^{eff}(I_m\rightarrow0)$ as a function of $I_m$.
Aiming at a comparison with  our experimental data,  we plot our frictional data expressed as the effective friction $\mu^{eff}(I)$ as a function of the inertial number $I$ in figure~\ref{fig_mueffIwetdry}a for our dry and wet samples.  
This plot tends to scatter the wet data in comparison with figure~\ref{fig_wet}b,  because they are not perfectly at the same $\mu_c$ as expected from equations~(\ref{mueffwet00}) and~(\ref{mueffwet0}). However, when normalized by  $\mu^{eff}(0)=\mu+\mu_c$, the wet data gather (figure not shown). 
 
 Our wet experiments are  realized at approximately the same  cohesion index $\xi\simeq 0.2 \pm 0.05$, whereas it is equal to $0$  for our dry samples; they have Stokes numbers $St\in [3\, 10^{-2}$-$2\,10^{-1}]$ for shear rates $\dot{\gamma}\in [5$-$35]$s$^{-1}$.  Relevant values $\alpha$ and $\beta$ for our materials are needed to compute $I_m$. 
 As a first guess, we use the values identified by the authors~\citep{Vo20}: $\alpha=0.062$ and $\beta=0.075$, but this leads for our experimental parameters to $I_m\simeq I$ so that this does not allow our data for the two systems studied here (wet and dry) to collapse.  
Some values of $\alpha$ and $\beta$ allow for a collapse of $\widetilde{ \mu^{eff}}(I_m)$, but their identification is not unique (and differ from several orders of magnitude), probably due to the small range of  explored values of $St$, $\xi$ and $I$ and to the limited precision of our data. For example, any value of $\alpha\in[10^{-3}$-$10]$ associated with any value of $\beta\in[5$-$10]$, but not limited,   allow a reasonable collapse. They lead to $I_m/I$ that is between $4$ and $20$, which value depends  on $St$ mainly and on $\xi$ slightly. Within our experimental uncertainties, even $I_m/I$ taken approximately as a constant allows for a collapse of our data, as shown in figure~\ref{fig_mueffIwetdry}b, where we plot  our experimental data of $\widetilde{ \mu^{eff}}$ for our dry and wet samples as a function of the approximation  $I_m\simeq 7 I$. 
The extended unified number $I_m$ proposed by~\citep{Vo20} is thus compatible with our experimental data, but the identification of the respective contributions of viscous, cohesive and inertial effects (i.e. the identification of $\alpha$ and $\beta$) is still an open question.

To finish, one can wonder where should be included the emergence of a cohesion stress (whether it is defined at the grain scale or at the macroscopic scale) in a granular material: in the relaxation contribution due to confining effects (added to the pressure, changing the extended inertial number) as done in~\citep{Vo20} and/or in the rheological law as a yield-stress as done  here with the Mohr-Coulomb criterion and/or in an additional friction coefficient~\citep{Vo20}. In others words, is it like an additional pressure and/or shear stress and/or friction coefficient?

\begin{figure}
	\centering
	 \includegraphics[width=.45\linewidth]{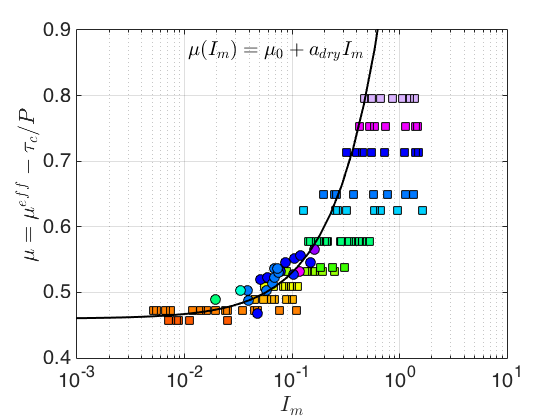} 
\caption{Wet and dry granular flows. Internal friction $\mu=\tau/P-\mu_c$  as a function of the modified inertial number $I_m$ defined by~\citet{Vo20}. Colors encode for the incline slope angle $\theta$ as in figure~\ref{fig_param}. 
}
	\label{fig_muIm}
\end{figure}

\section{Conclusion}
\label{conclu}
In this paper, we have been concerned with both theoretical and experimental aspects of the flow of unsaturated wet cohesive granular materials down rough inclined 
planes. Understanding the constitutive relation of cohesive granular materials is clearly of fundamental importance to the study of granular flows in a geophysical context as well as to the handling of granular materials in industry where small amounts of a wetting fluid create 
cohesive contacts between the grains. We have shown that the introduction of a constant cohesion in the rheology described by equation~(\ref{eq_rheocohesionstat}), namely the Mohr Coulomb yield criterion leads to a modified flow 
typology compared to the cohesionless case: a plug flow is then observed under the free surface. Indeed, cohesive granular materials generically form two different flow 
regions: a 
sheared region topped with a solid region (the plug) extending below the free surface characterized by a vanishing shear rate of strain.
Theoretical analyses have shown that the width of this plug region scales like the cohesive length $h_c$, given by equation~(\ref{eq_hc}) which is independent on total thickness of the granular layer but dependent on the tilt angle $\theta$, the cohesive stress, the mass density of the wet granular material; it diverges when $\theta$ approaches the angle of repose $\arctan(\mu_0)$. We have also found that in the flowing region below the plug, the  
velocity profile has two contributions of opposite signs: one   similar to the Bagnold velocity profile (power $3/2$) and the other one induced by cohesion (power $1/2$).

Concerning the rheology, we have shown that the extension of Mohr Coulomb law to the inertial regime gives predictions in quantitative agreement with experimental measurements only when one considers  that dry and wet samples have straightforwardly different internal friction commonly described by the so-called $\mu(I)$-rheology.

Notice that we analysed our experimental data in terms of the Mohr-Coulomb yield criterion and we proposed to apply the knowledge  acquired on friction  in dry granular flows to the internal friction  $\mu(I)$. In particular, we used the  empirical known scalings (affine relation or more elaborate function) to describe the internal friction  $\mu(I)$. This is also what is done by~\citep{Abramian20} in their recent numerical study. Alternatively, some authors used the  empirical known scalings to describe the effective friction $\mu^{eff}(I)$~\citep{Vo20}. But this would lead for our wet sample to a non constant cohesion stress in the framework of the Mohr-Coulomb yield criterion, as if the cohesion stress was not intrinsic to the material, but dependent on the loading conditions. The same would happen if we express our rheological law for wet flows from the one for dry ones~\citep{Badetti18} with $\tau = \mu_{dry} P + C$: this leads to a cohesion $C$ that depends on the pressure, so that it is not a material's property. In general, whereas the identification of the internal friction coefficient  $\mu$ (instead of the effective one $\mu^{eff}$)
will allow for predictions by incorporating the constitutive law  $\tau = \tau_c + \mu P$ in motion equations in the framework of continuum mechanics, the effective friction coefficient $\mu^{eff}$ may be useful only for  steady flows. A future perspective could be to check whether the  extended unified visco-cohesive-inertial number $I_m$ proposed by~\citet{Vo20} allows to collapse the internal friction coefficient  $\mu(I_m)$ to be compatible with the Mohr Coulomb yield criterion. For the two systems studied here, $I_m$ can be defined to collapse our experimental measurements of the internal friction coefficient  $\mu(I_m)$, as seen in figure~\ref{fig_muIm}.   

Additionally, it would be helpful to investigate the effect of the liquid content as well as the rheological effects of an increase in the viscosity and surface tension of the wetting liquid, in order to identify the  relevant rheological laws accounting for capillary and viscous forces, in addition to inertial effects.

\vspace{1cm}

Acknowledgments. We would like to acknowledge the students Périg Le Jeannic, Khady Seck, Emma Charanton, Samy Idrissi-Kaitouni and Mathurin Romand. We also thank David Hautemayou, Cédric Mézière, Christophe Courrier, Thierry Bastien and Jean-Marie Citerne for technical helps for the experimental set-up. A. Fall thanks the Agence Nationale de la Recherche (Grant No. ANR‐16‐CE08‐0005‐01) for funding.  We thank Lydie Staron for her suggestions on the manuscript.

\vspace{1cm}  

  Declaration of interests. The authors report no conflict of interests.

\bibliographystyle{jfm}
\bibliography{WetGranu_biblio}

\end{document}